# Differential aperture photometry and digital coronagraphy with PRAIA

M. Assafin[a,b]

[a]*Universidade Federal do Rio de Janeiro - Observatório do Valongo, Ladeira do Pedro Antônio 43, Rio de Janeiro, 20.080-090, RJ, Brazil*
[b]*Laboratório Interinstitucional de e-Astronomia (LIneA), Rua General José Cristino 77, Rio de Janeiro, 20.921-400, RJ, Brazil*

**Abstract**

PRAIA – Package for the Reduction of Astronomical Images Automatically – is a suite of photometric and astrometric tasks designed to cope with huge amounts of heterogeneous observations with fast processing, no human intervention, minimum parametrization and yet maximum possible accuracy and precision. It is the main tool used to analyse astronomical observations by an international collaboration involving Brazilian, French and Spanish researchers under the Lucky Star umbrella for Solar System studies. Here, we focus on the concepts of differential aperture photometry and digital coronagraphy underneath PRAIA, used in the reduction of stellar occultations, rotational light curves, mutual phenomena and natural satellite observations. We highlight novelties developed by us and never before reported in the literature, which significantly enhance the precision and automation of photometry and digital coronagraphy, such as: a) PRAIA's pixelized aperture photometry (PCAP), which avoids pixel sub-sampling or fractioning; b) fully automatic object detection and aperture determination (BOIA), which abolishes the use of arbitrary sky background sigma factors, and finds better apertures than by using subjective FWHM factors; c) better astrometry improving the aperture and coronagraphy centre, including the new Photogravity Center Method besides circular and elliptical Gaussian and Lorentzian generalized profiles; d) coronagraphy of faint objects close to bright ones and vice-versa; e) use of elliptical rings for the coronagraphy of elongated profiles; f) refined quartile ring statistics; g) multiprocessing image capabilities for faster computation speed. We give examples showing the photometry performance, discuss the advantages of PRAIA over other popular packages for Solar System differential photometric observations, point out the uniqueness of its digital coronagraphy in comparison with other coronagraphy tools and methods, and comment about future planed implementations. Besides Solar System works, PRAIA can also be used in the differential photometry of variable and cataclysmic stars and transient phenomena like exoplanet transits and microlensing, and in the digital coronagraphy of astrophysical observations. PRAIA codes and input files are publicly available for the first time at: https://ov.ufrj.br/en/PRAIA/.

## 1. Introduction

Digital coronagraphy subtracts the brightness profile of a source. The "true" brightness profile of the target is then recovered or dramatically enhanced in the image. It is very important for astrometric and photometric measurements (Assafin et al., 2009; Camargo et al., 2015; Dias-Oliveira et al., 2015).

Photometry is essential for rotational light curves, stellar occultations and mutual phenomena. Mutual phenomena allow for milliarcsecond astrometry of planetary satellites (Morgado et al., 2019), and can be used to estimate precise sizes and shapes for binary asteroids (Descamps et al., 2007). Rotational light curves coupled with stellar occultations allow for the precise determination of the sizes and shapes of asteroids and TNOs with km accuracy (Santos-Sanz et al., 2022). Stellar occultations allow the search for putative atmospheres in TNOs, dwarf planets and natural satellites, and detection of faint rings, moonlets and disperse material around minor bodies (Morgado et al., 2023).

Since the 90's, a team of Brazilian researchers known as the Rio Group has led astrometric works in reference systems and Solar System bodies (Assafin 2023 and references therein). Since the mid-2000's, it publishes many photometric studies of Solar System bodies by using stellar occultations and mutual phenomena (see Table 1 in Section 6.1). The Rio Group is in an international collaboration under the Lucky Star Project[1] with other two French and Spanish teams. Lucky Star uses many small (40 – 80 cm) to mid (1 – 2 m) to large (4 – 8 m) aperture telescopes in Chile, Hawaii, Spain, France, Brazil, Australia and Africa countries with many types of CCD detectors. We also maintain a fruitful interaction with the amateur community around the world, which actively participates on our stellar occultation campaigns. Amateurs are extremely important, allowing for the coverage of many events that would otherwise be missed. However, telescope apertures range from only 20 cm to 1 m, and observations may be done under non-ideal sky/technical conditions with poor guiding, low signal-to-noise ratios (S/N), few calibration stars, timing issues, etc.

Although stellar occultations and mutual phenomena last only a few minutes, the amount of acquired images is huge (5 – 10 GB or more) due to the very short time exposures (usually less than 1 s), necessary for improving the time/spatial resolution. This and the heterogeneity of the data from many telescope types, detectors, sky conditions, S/N regimes, made us soon realize that the use of standard stellar photometry packages such as the Image Reduction and Analysis Facility (`IRAF`[2], Tody 1986, 1993) and `DAOPHOT` (Stetson, 1987) was not practical due to the excessive parameterization and human interaction.

---

[1]Lucky Star: https://lesia.obspm.fr/lucky-star/index.php

The solution was to develop our own pipeline of differential aperture photometry – the `PRAIA` photometry task (`PPT`). As only relative flux is important, differential photometry was a clear choice. Here, aperture photometry clearly renders better results than Point Spread Function (PSF) photometry, because characterizing the ever-varying PSF of the target in a stellar occultation or mutual phenomena is problematic. The involved objects may not be punctual light sources. Even if they were, the target resulting from the mix of objects will not. Frequently, a severe or total flux drop occurs in eclipses and stellar occultations, making it impossible to set and measure a PSF.

One of the main difficulties in the photometry and astrometry is the contamination of a target by the surrounding scattered light of a nearby source in observations made without coronagraphs. It soon became clear that removing the scattered light by the direct approach of fitting and subtracting PSF profiles did not render the best results. Another issue was bad guiding, resulting on elongated profiles difficult to model. Because no package was available in the 90's or 2000's (and even up to date) to cope with these problems and with our voluminous and heterogeneous sets of observations, we developed our own pipeline from the start, and so the `PRAIA` Digital Coronagraphy task (`PDC`) was born. Best results could then be obtained with `PDC` by using other forms of manipulating the pixel counts of the images.

Here, we describe `PRAIA`'s photometry and coronagraphy. In Section 2, we present the package, evolution and current concepts of the differential aperture photometry and digital coronagraphy tasks. We also highlight `BOIA` (Browsing Objects: Identification and Analysis), a new robust automatic precise way of detecting objects, defining the best aperture and computing their image properties for photometry and coronagraphy procedures. In Section 3, we describe our implemented methods of differential aperture photometry. There, we introduce a new photometry concept, Pixelized Circular Aperture Photometry (PCAP). Section 4 deals with the building of light curves from photometry measurements, and the computation/removal of intruding flux $\phi_o$. In Section 5, we introduce our digital coronagraphy procedures, described in detail in Appendix A. Section 6 presents all publications that used `PRAIA` photometry between 2006–2022, certifying it in practice. We give examples attesting PPT's performance and make comparisons with `DAOPHOT` photometry. We then discuss the advantages of `PRAIA` photometry over other available packages in the context of our Solar System photometric work. Section 7 validates `PRAIA` coronagraphy by commenting some examples, pointing out published works that used it, and by comparing it with other algorithms published in the literature. Conclusions and remarks about future photometry and coronagraphy updates are drawn in Section 8.

## 2. PRAIA – Package for the Reduction of Astronomical Images Automatically

`PRAIA` stands for "Package for the Reduction of Astronomical Images Automatically". It is a standalone package developed in `FORTRAN` with no graphic interfaces composed of multiple independent tasks. Each task has a source/executable code and an input text file (ASCII format) with editable input parameters. The tasks can be run in scripts for the astrometry and photometry of observations in (standard or not) Flexible Image Transport System (FITS, Wells et al. 1981) format. `PRAIA` allows for full manipulation of input/output data with user's own preferred external statistical, plotting and visualization tools. Due to the large amount and heterogeneity of observations in our works, we developed `PRAIA` with full automation, minimum parameterization, fast processing speed and precision. `PRAIA` is certified by its use in many works published in prestigious journals (Assafin 2023; see also Table 1 and 2). Detailed installation/usage instructions for `PRAIA` tasks with codes, input files and User Guide documentation are for the first time publicly available[3].

### 2.1. The PRAIA photometry task (PPT)

The aim of the `PRAIA` photometric task (hereafter `PPT`) is twofold: do differential aperture photometry over digitized FITS images and build up light curves. The task also allows for the computation of intruding fluxes in light curves. `PPT` is suited for (but not limited to) the treatment of differential photometric observations of Solar System bodies (asteroids, Centaurs and TNOs, natural satellites, planets and dwarf-planets) for the analysis of rotation, stellar occultations and mutual phenomena.

`PPT` is in its fourth version. It underwent three major upgrades. The first version 0 (2006–2008) had many implementations for automatic processing, version 1 (2009–2012) introduced non-integer radius apertures, and version 2 (2013–2018) improved the tracking of moving objects. A complete historical review of all previous versions is given in the User Guide.

Version 3 (since 2019) underwent major improvements and is the current version in use. We introduced a bisection method for automatically determining the best set of fixed apertures for the objects. Variable apertures are allowed with a variety of options to circumvent practical problems. A new centring algorithm suited for faint objects was developed – the Photogravity Center Method (`PGM`). We also implemented the novel Pixelized Circular Aperture Photometry (PCAP). It copes nicely with area preservation and flux equalization on fixed and varying aperture photometry, dispensing cumbersome and precision-limited fractional pixel computations of circular areas, with gain in precision and processing time. Another novelty developed was `BOIA` (Browsing Objects: Identification and Analysis) – a fully automated way to detect objects and find optimum aperture sizes, by computing the local sky background and setting aperture limits with Student-t and F-tests and quartile statistics. This avoids the common, subjective and usually imprecise practice of setting up threshold values by fixing sky factors. The new internal manipulation of pixels and memory usage also significantly improved the processing speed. A new light curve mode was also developed for computing the intruding flux of a body, useful in the analysis of stellar occultations by bodies with atmospheres.

---

[2]`IRAF` was distributed by the National Optical Astronomy Observatory, which is operated by the Association of Universities for Research in Astronomy (AURA) under cooperative agreement with the National Science Foundation. Currently it is supported by the IRAF Community: https://iraf-community.github.io

[3]`PRAIA` is publicly available at https://ov.ufrj.br/en/PRAIA/

PPT makes photometry measurements in four modes (modes 0, 1, 2 and 3), and builds light curves from photometry measurements or other light curves in two modes (modes 4 and 5). Fixed aperture photometry is made with modes 0 (manual entries) and 1 (bisection method for automatically finding best fixed apertures for all objects). Varying aperture photometry is made with modes 2 (manual object indication) and 3 (semi and fully automatic object identification). Mode 4 builds light curves from photometry measurements. Mode 5 computes the intruding body flux $\phi_o$ in stellar occultations from other light curves, also producing auxiliary light curves in the process.

*2.2. PRAIA Digital Coronagraphy (PDC)*

The PRAIA digital coronagraphy task (hereafter PDC) removes the brightness profile of an object - the Source. Usually – not necessarily – the Source is a bright object. The final goal is to enhance or ideally recover the brightness profile of a target, which is affected by the contaminating scattered light of the nearby Source. PDC is suited for photometric and astrometric Solar System works involving natural satellites close to bright planets, and stellar occultation target stars close to field stars. PDC also serves for astrophysical studies as well.

PDC is in its fourth version. It had three major upgrades. In its first version 0 (1995–2007), removal of the scattered light of the Source was done by the count subtraction of pixels symmetrically located around a straight line crossing the Source (see Figures 1 and 2 in Veiga & Vieira Martins 1995). Version 1 (2007–2008) introduced the use of circular rings for the computation of the source profile (Assafin et al., 2008), and version 2 (2009–2019) improved ring statistics. The User Guide gives a complete historical review of all previous versions. In all these versions, the Source had to be the brightest object in the FOV (Field-Of-View) working area. Version 3 (since 2019) underwent major improvements in all aspects with respect to the previous versions, including better centring, statistics and the use of elliptical rings. It is described in this paper.

In the current PDC version (3), the sky background of the entire FOV is first flattened. We now work with elliptical rings to account for non-rounded images. Besides the Source, we introduced a new class of object: the Reference Object. The Reference Object may be useful both in the determination of the coronagraphy centre as in the setup of elliptical ring's parameters. In particular, the Reference Object now allows for the coronagraphy of Sources not only fainter than targets, but also even closer to them too.

The current coronagraphy procedures are still divided in two steps: the determination of the coronagraphy centre and execution of the coronagraphy itself. The $(x_C, y_C)$ Source's coronagraphy centre can now be directly input or set relative to the coordinates of the Reference Object in a variety of convenient ways. The coronagraphy algorithm itself also changed: now we use elliptical rings and quartile statistics instead of histograms in the computation of the Source profile. BOIA was also adapted to deal with elliptical shapes, being used to compute elliptical parameters and set apertures for object centring. In this paper, we describe all coronagraphy procedures present in PDC.

*2.3. BOIA: object detection, aperture, properties, centre*

Browsing Objects: Identification and Analysis (BOIA) unites original procedures developed by us, aiming for the automatic detection, aperture setup and characterization of the image properties of objects in the FOV. It is present in parts or in totality in all photometric and astrometric PRAIA tasks. A complete description of BOIA used in its full form is in Assafin (2023).

Photometry (mode 3; guide in all modes – see Section 3.5) and coronagraphy (Reference Object) have objects automatically detected by scanning valid FOV windows in the images by 3x3 pixel cells and ordering by total flux. From highest to smallest fluxes, optimized apertures and sky background rings are searched and determined around the initial $(x_i, y_i)$ cell centres. The search stops when aperture fluxes match those of sky background rings, as indicated by Student-t and F tests of significantly different means (Press et al., 1982). Superposed detections are dropped, except the largest aperture. Initial $(x_i, y_i)$s are available by default in manual/semi-automatic photometry modes 0, 1, 2, and for the Source object in coronagraphy.

Optimized found apertures and sky background rings have the highest S/N – BOIA's main principle. The same procedures are applied for all objects – see a complete description in Assafin 2023). Successive rings are sampled around the initial $(x_i, y_i)$, quartile statistics filter discrepant counts, and we apply Student-t and F tests of significantly different means (Press et al., 1982) to automatically detect the sky background without sky background sigma factors or other subjective threshold settings. The initial $(x_i, y_i)$ is updated until we properly engulf the object. Values for the aperture size and sky background ring size and width are sampled. In photometry, an alternative 2D circular Gaussian centre is also tested. Apertures and rings are circular in photometry, and elliptical in coronagraphy when the semi-major axes are sampled (instead of radii in the circular case), maintaining the eccentricity and the orientation angle. In fixed aperture photometry, only the aperture centre is tested, while for varying apertures all aperture/ring parameters are sampled, including the centre. BOIA keeps the aperture and sky background ring set with the best S/N.

We have shown (see Section 10.2.1 in Assafin 2023) that BOIA explains and expands the old astrometry/photometry axiom that recommends the use of 2–3 Gaussian $\sigma$ radius apertures. A natural relation emerges between measured BOIA aperture radii and the FWHMs, not only confirming the axiom but refining the connection between the aperture size and the FWHM in a case by case basis, inside and also outside the 2–3 $\sigma$ limits toward both brighter and fainter objects. This means that aperture sizes found by BOIA are better than arbitrarily setting the radius as 2–3 $\sigma$, or using other empirical relations with the FWHM.

BOIA's main astrometric/geometric object properties are derived after the best aperture is found – see Section 3.6 in Assafin (2023) for details. BOIA's centre, semi-major axis $a$, semi-minor axis $b$ and orientation angle $\theta$ are computed by using the high S/N pixels inside the optimum aperture. In coronagraphy, during the search for the best elliptical aperture/sky background ring set, the semi-major axes, eccentricity and the orientation angle are computed in the same way. In Section Appendix A.5 we show how to know if pixels are inside an elliptical aperture or ring.

In photometry and coronagraphy, on the best aperture/ring search, $(x, y)$ centres are updated with the Photogravity Center Method (PGM, see Assafin 2023). PGM presents a novel kind of moment centring algorithm, which also takes into consideration pixel clustering, being superior to the largely used Modified Moment (Stone, 1989), mostly for faint objects. Unlike moment methods, PGM also allows for the estimation of $(x, y)$ errors.

Besides PGM, precise photometry aperture centring can be done with 2D circular Gaussian PSFs (Point-Spread-Function). Object coronagraphy centring also includes 2D elliptical Gaussian profiles, and circular and elliptical extended Moffat profiles (Moffat, 1969), i.e. general Lorentzian PSFs with free-fitted $\alpha$ and $\beta$ coefficients – see details in Assafin (2023).

## 3. Differential aperture photometry with PRAIA

In many Solar System and astrophysical transient phenomena, we only need the relative variation of fluxes rather than measuring absolute fluxes in standard photometric systems. If observations are made with a sufficiently small FOV, one can profit from the differential photometry technique and measure relative flux variations with high internal accuracy and precision. In differential photometry, non-photometric-standard stars or even Solar System bodies can be used as photometric calibrators. Since the same systematic effects affect the target and calibrator fluxes in small FOVs, systematic errors are canceled out after the calibration procedure, resulting in high-precision, highly accurate relative flux measurements. No extra calibration images are needed, simplifying telescope observations and data treatment without any concession to accuracy/precision.

Differential photometry is successfully applied in rotational light curves, stellar occultations and mutual phenomena. Rotational light curves are used in the determination of axis proportions and eventually in the density limits of bodies. Stellar occultations serve to obtain highly accurate/precise sizes and shapes for the occulting bodies, mostly if combined with rotational data. Mutual phenomena furnish highly accurate/precise relative positions in the sky plane between the involved bodies.

Due to the nature of stellar occultations and mutual phenomena observations, we avoid Point-Spread-Function (PSF) photometry. In stellar occultations the target is made of two mixed PSFs – those of the body and of the star to be occulted. The apparent distance between these objects vary along the event and at some point the body may only partially obstruct the star. A similar situation arises in mutual phenomena with two bodies at play. The resulting PSF model in these cases is unclear and the use of Gaussian or Moffat profiles may yield to biased results. Even if an empirical PSF can be numerically obtained, the procedure is highly computer-time consuming, with no guarantee of satisfactory results. In stellar occultations, short exposure times let objects in a low signal-to-noise ratio regime for better time/spatial resolution, and the target star may be totally occulted, making it very difficult or impossible to fit a PSF.

For these reasons, PPT only uses aperture photometry. Aperture photometry techniques are more flexible and can successfully measure the light of targets in all situations described above, even in the case of partial or total flux drop such as in stellar occultations and mutual phenomena.

### 3.1. Flux, sky background, S/N and flux errors

PPT sums ADU (Analogical Digital Units) counts of all pixels within the aperture (actually, this procedure is more complex, see Section 3.3. This raw flux is subtracted from the contaminating sky background flux. The sky background flux is computed from the ADU counts of a given circular sky background ring. Once the ADU value of the sky background contribution per pixel is determined, one simply subtracts the raw flux from the overall sky background contribution of all pixels within the aperture to derive the "clean" aperture flux of the object.

PPT uses quartile statistics as a more robust and precise way of computing the sky background than with histograms. The 25 percent lower and higher pixel counts within the sky background ring are eliminated, avoiding the eventual presence of cold pixels and nearby contaminating objects. The remaining pixels are used to obtain the average and standard deviation of the sky background. Histograms can get ill-sampled yielding to bad statistics, specially when few ring pixels are available.

The signal-to-noise ratio (S/N) is a fairly important quantity for PPT. It is a function of CCD parameters and photometry measurements of the object and sky background, as in the classic formula by Eq. 1, where $N$ is the total number of photons received at the object and subtracted from the sky photons, $n_{pix}$ is the number of pixels associated to the object, $N_s$ is the number of sky photons per pixel, $N_d$ is the photon dark current per pixel and $N_r$ is the CCD read noise in electrons per pixel (Howell, 1989). The conversion between CCD electrons and photons come from the quantum efficiency of the CCD per wavelength bin. The conversion between CCD electrons and ADU units comes from the gain factor $g$ in electrons per ADU unit.

$$\frac{S}{N} = \frac{N}{\sqrt{N + n_{pix}(N_s + N_d + N_r^2)}} \quad (1)$$

Maintaining its exact meaning, Newberry (1991) rewrites Eq. 1 in a more practical form only in terms of ADU counts, as in Eq. 2, where $C_0$ (in ADU units) is the clean object flux, $g$ is the gain, $n$ is the number of object (aperture) pixels, $\sigma_{bg}$ is the sky background dispersion (in ADU units) and $p$ is the number of ring pixels used in sky background computations.

$$\frac{S}{N} = \frac{\sqrt{C_0}}{\sqrt{\frac{1}{g} + n\,\sigma_{bg}\,C_0^{-1}(1 + \frac{1}{p})}} \qquad (2)$$

PPT allows for the use of both "classical" (Eq. 1) and "ADU-based" (Eq. 2) formulas for computing the S/N of the objects.

From the S/N, the flux error $\sigma(F)$ in ADU units, the (dimensionless) relative error $\sigma(F)/F$ and the relative error $\sigma(m)$ in magnitudes are expressed in Eq. 3:

$$\sigma(F) = \frac{F}{S/N} \; , \; \frac{\sigma(F)}{F} = \frac{1}{S/N} \; , \; \sigma(m) = \frac{2.5/ln(10)}{S/N} \qquad (3)$$

*3.2. Fixed and varying aperture sizes*

Fixed apertures and sky background rings are used when the sky condition is steady along the series of images. The best fixed aperture/ring sizes should be chosen after evaluating a group of images for which the objects would be known not to vary their true fluxes. The best set should correspond to the results with the least flux dispersion with time. The use of different but fixed aperture sizes for each object is ok, since we are only interested in measuring the relative variation of fluxes.

Varying apertures are used when the sky transparency, atmosphere turbulence and instrument are an issue. This is often the case of transient phenomena such as stellar occultations or mutual phenomena, which must be observed at specific dates and places regardless of the local sky conditions. Besides, it often implies in the participation of smaller observatories and/or the amateur community, with the use of less robust smaller aperture telescopes and poorer detectors.

PPT can optimize both aperture kinds. For fixed apertures, it can search for the best set of apertures and sky background rings by brute force or using a bisection method, aiming at the least flux dispersion outside transient events. For varying apertures, the determination of the best set for each object in each image of a time series is governed by the S/N of the objects.

Varying apertures involve the compensation for the loss/gain of relative flux from one object to the other due to the use of different aperture sizes. This flux/area equalization is done with light growth curves made from the flux of bright objects. These curves are traditionally expressed as a function of aperture radius. PPT goes more straight to the point by making light growth curves tied to the aperture *area* in pixel units – a photometry novelty to the best of our knowledge.

In two cases fixed apertures may give equivalent/better results than using varying apertures. First, if the brightest object used to equalize fluxes/areas is not bright enough, the flux noise added in the process may render worse results than choosing good fixed apertures. Second, if the target and/or calibrators are near another objects or near the FOV borders in some, most or all images, then the optimum aperture may fall outside the FOV, or be contaminated. PPT allows for setting size limits in varying aperture photometry, but in some cases better/unbiased results are obtained with fixed smaller apertures.

Many aperture photometry options are available by PPT: different, but fixed apertures for each object in all images, and variable apertures with flux/area equalization for each or all images. These flexible options together with other functionalities allow us for implementing the best photometry strategy for the observations in a case by case basis.

*3.3. Pixelized Circular Aperture Photometry (PCAP)*

Care must be taken with the circular (or any non-rectangular) shape of apertures due to the square form of pixels encompassed by object profiles in CCD or digitized images. It is computer-time expensive to calculate for all measurements the partial area of pixels intersected by the aperture circumference, be it made numerically by pixel sub-sampling or with analytical formulae linked to complicated geometrical schemes – see for instance Gendler et al. (2018). Besides, there are no satisfactory extrapolation schemes that can precisely enough furnish the true flux under a given partial pixel area, unavoidably resulting in a net flux error propagation in the measurements.

Bickerton & Lupton (2013) give a good summary of this known and unsolved problem in classical circular aperture photometry work. Nevertheless, they present a way to measure the "exact" flux within the exact circular area of a given circular aperture, under few assumptions, by the use of `sinc` functions. The procedure involves Fast Fourrier Transformation computations and predefined indexing of parameters for a wide range of aperture radii. The method can be extended to elliptical apertures and will be applied on LSST aperture photometry. Unfortunately, their method involves heavy computations at some stages, being only suited for a single and quite stable instrument/detector, as is the LSST case. This condition is opposed to the variety of instruments/detectors used in the photometric observations within our collaboration around the world.

Mostly for small apertures, given circle apertures with the same nominal radius, the number of pixels inside the circle vary as the aperture centre lays on distinct parts of the central pixel – a pixel-centring effect that occurs in mask-based photometry. Depending on centring, even using the same nominal aperture radius, the number of internal pixels contained inside the circular aperture could vary, even for the same object from image to image, as the object centre varies even slightly in the FOV.

Thus, even when using fixed apertures, and certainly in the equalization of flux/area with variable apertures, if we take pixels internal to a circular aperture to measure flux, we still may be unwarily sampling different numbers of pixels for the same nominal aperture radius, i. e. taking different areas, and thus different fluxes. If we do not take this into account, we are ultimately adding noise to the measurement of the variation of the flux with time. It is not uncommon to find targets in rotation light curves, stellar occultations or mutual phenomena with small circular aperture sizes, when the problem gets even worse.

All this motivated us into a new circular aperture photometry approach – the Pixelized Circular Aperture Photometry (PCAP), distinct from pixel mask photometry. PCAP uses circular aperture radii only in a nominal basis, and does not try to compute or make extrapolations to obtain the flux that would be encompassed by an exact circular area. The photometric principle is to only measure an integer number of pixels which corresponds as close as possible to the area of a circular aperture with a given nominal radius, thus avoiding any kind of pixel sub-sampling, intersection area calculations or flux extrapolation schemes.

PCAP does not parameterize the photometry by the nominal radius $R$ of the aperture, nor by its nominal circular area. It parameterises the flux with respect to the *area* defined by an integer number of pixels that corresponds to the round off of the nominal area of the circular aperture of nominal radius $R$. Take for example an aperture with nominal radius $R$=3.3 pixels[4]. Now taking 1 pixel as the unit for area (the area encompassed by 1 pixel), the total nominal area is $TA = \pi R^2 = 34.212$ pixels, while the round off area is $ROA = 34$ pixels. Thus, the round off area $ROA$ associated to an aperture of radius $R$=3.3 pixels to be used in PCAP is $ROA = N = 34$ pixels, where $N$ is the number of actual pixels to be used in the aperture flux measurements. The pixels are sorted in accord to their distance to the object's centre. This properly takes into account the pixel-centring effects of mask-based photometry – for any given nominal aperture radius value $R$, there is always an unique number $N$ of pixels associated to it, from the round off of the total aperture area in pixel units. PCAP can also be easily adapted for elliptical or irregular aperture shapes.

### 3.4. Flux/area equalization for fixed and varying apertures

PCAP allows for a robust equalization of the fluxes and areas of the apertures. The equalization is made in two ways depending whether we are dealing with fixed or varying apertures.

For fixed apertures, PCAP equalizes areas straightforwardly. Each object $i$ is measured on each image with its fixed respective aperture size $R_i$ with associated pixel number (area) $N_i$. The sampled pixels are sorted in crescent distance order from the object centre, so that the resulting $N_i$ pixels always covers the same area from image to image, but also adapts its form to essentially cover the same object's portions in the images, regardless of pixel-centring effects. In this way, by force, we are always measuring flux over exactly the same effective area around the object centre, for each object in all images.

For varying apertures, PCAP computes the flux $F_i$ of each object $i$ from its optimum BOIA aperture (see Section 2.3), and register the nominal radius $R_i$ and associated number of pixels $N_i$. The aperture radius $R_i$ may vary considerably from object to object, and from image to image. PCAP equalizes both fluxes and areas simultaneously by constructing a Light Growth Curve (LGC), as described next (see Howell 1989; Stetson 1990).

If not imposed, PCAP arbitrarily fixes the aperture radius $R$ of the faintest target in the reference image as the reference aperture, which corresponds to the reference number $N$ of pixels related to the round off area of this aperture. Then, for each image, PCAP picks up the brightest object in the FOV and constructs the observed LGC by sorting pixels in crescent distance from the centre and accumulating the flux as a function of the number of pixels (area). For each object $i$ of each image, we use the observed LGC of that image to deduce the ratio $F(N)/F_i(N_i)$, which represents the flux correction factor to take $F_i$ to the corrected flux $F$ of the object for an equalized area of $N$ pixels. Since the accumulation of flux grows smoothly only for pixels within the same distance from the centre, we only use the pixels of the observed LGC that are within a ring of 1 pixel width and same radius as that of the object's aperture. Since we want to use as much information as possible, instead of computing the factor $F(N)/F_i(N_i)$ from $F(N)$ and $F_i(N_i)$ directly available from the observed LGC, we fit the observed LGC by a polynomial $P$ and get $F(N)/F_i(N_i)$ from $P(N)/P_i(N_i)$. Empirical tests showed that a polynomial $P$ of degree 2 fits well any observed LGC, when we take $F$ as a function of the the number of pixels $N_p$. Fig. 1 displays an example of a polynomial fit to an observed LGC, where we display the number of pixels $N_p$ used in the fitting. The (O-C) residuals (LGC – $P$) is also plotted as a percentage of the LGC. The equivalent radius shown in the residual's plot is the radius of a circle with the same area as the number of pixels $N_p$. The plots indicate that $P$ represents the observed LGC within less than 0.06 per cent.

---

[4]Here, "pixel" has the usual meaning "length of the sides of a square pixel"

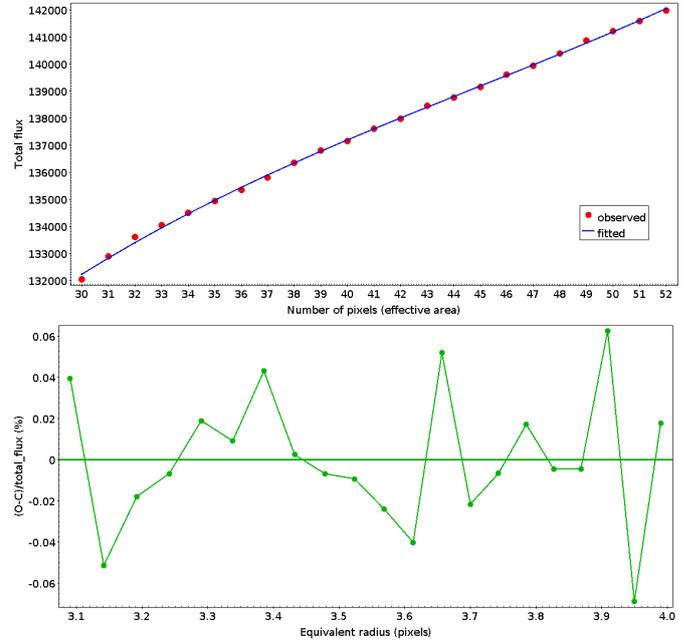

Figure 1: Top panel: Observed Light Growth Curve (LGC) (red) with accumulated pixel flux as a function of the number $N$ of pixels sorted by distance from the centre of the brightest object on a FOV; a polynomial $P$ of degree 2 (blue) was fitted as a function of $N$. Lower panel: (O-C) residuals = LGC – $P$ as a percentage of LGC; the equivalent radius in the plot is the radius of a circle with the same area as the number of pixels in the top panel. The polynomial model represents well the LGC within less than 0.06 per cent.

LGCs are evaluated for each object in the image to find the flux factor that will equalize each area (number of pixels) to the reference area. PCAP allows us to set the reference aperture radius $R$ for each object. This is equivalent to setting fixed apertures for each object for all images, but with the benefits of the varying aperture photometry. Unlike apertures, the sky background level and dispersion are completely insensitive to flux correction issues. Internally, PPT works directly with sky background level and dispersion in ADU/pixel units, so that sky background corrections can be easily applied when dealing with flux/pixel units, as is the case with PCAP. Also, for saving computer time, PPT only works internally with integer pixel units for the sky background ring radius and width.

PCAP proved to be more precise and faster than the method of pixel sub-sampling and/or internal/external ring extrapolation procedures used in previous task versions. This is because no interpolation or extrapolation for computing partial ADU flux counts of partial pixels is done – only entire integer number of pixels are considered and manipulated straightforwardly.

We also do not use fixed masks anymore – masks with a fixed geometry layout form with a fixed area or number of pixels. Fixed masks force the sampling of inappropriate border pixels with distinct flux levels from image to image, depending on the location of the object centre within the central pixel, resulting on higher flux dispersion along the images. In PCAP, the pixels are naturally added by their distance from the object centre, and do not form a fixed mask with a predefined geometric format. The object is sampled from one image to the other by the same number of pixels, distributed as homogeneously as possible around the object centre, resulting on masks of variable geometric formats with the same fixed number of pixels that sample more coherent flux levels from image to image, deriving measurements with smaller flux dispersion.

*3.5. Guide, calibrators, targets and ghosts, the reference image*

PPT works with 4 categories of objects: the guide, calibrators, targets and associated ghosts. The guide serves to automatically track calibrators and targets in the FOV of all images. Calibrators serve to calibrate the flux of targets. Targets are the objects of scientific interest. A ghost is an "empty-object" centred at a fixed relative distance close to a target, sampling a region dominated by the local sky background, with the same aperture and sky background ring sizes of the target. Ghosts serve for sky background comparisons with their associated targets, and should present fluxes varying around zero, since it is the clean flux of the local sky background. A target or a calibrator may also be the guide object. The guide, the calibrators and the targets may be fixed or moving objects in the FOV.

Sometimes the FITS headers of images carry wrong data or lack sufficient information for the identification of objects by their right ascension and declination, even when the FOV contains enough reference stars. The FOV may also not contain enough reference stars, preventing precise pixel reduction to spherical coordinates even with some valid Word Coordinate System (WCS) keys. That is why PPT does not identify objects in the FOV by their right ascension and declination. In fact, PPT identifies calibrators and targets in all images regardless of any WCS/positioning header keys, making it quite a versatile tool.

PPT tracks calibrators and targets by their relative distances to the guide. Ghosts are located by their fixed distances to their respective targets. The guide must be the brightest object in a fixed region of the FOV for all images. This region can have any size. The guide is automatically identified in all images. For the correct location of calibrators, targets and ghosts, it is mandatory that: a) the same guide be always present within the furnished fixed region in all images; b) the calibrators, targets and ghosts be always within the FOV in all images; c) the FOV have the (x,y) axis orientation and pixel scale fixed or varying in a way that can be modeled by a polynomial of degree $N$ (typically $N \leq 2$) with time. These conditions are easily satisfied once images are acquired with the same instrument setup.

The initial pixel coordinates of calibrators, targets and ghosts, and then their initial relative distances to the guide are setup in the FOV of a *reference image*. These initial relative coordinates are then used to track the objects in the remaining images. The reference image should be the best sampled of the set with the highest S/N values for the guide, calibrators and specially for the targets. In PPT photometry modes 0 to 2 we select the reference image and indicates the initial pixel coordinates of calibrators, targets and ghosts (the guide too in mode 0). In photometry mode 3, the reference image and initial coordinates of targets may still be indicated, with ghost coordinates automatically computed. In full automatic form, mode 3 picks up the reference image by searching the guide with the highest S/N among all observations, then automatically detects the best calibrators, the targets and their ghosts in the reference image (Section 2.3), and computes their initial relative coordinates to the guide. From the initial relative distances, the tracking of objects is done for all images.

In the fully automatic detection of calibrators and targets of photometry mode 3 (see Section 2.3), all objects get provisional aperture sizes from the reference image, and are evaluated in all images. Objects with apertures falling outside the FOV are excluded. Only the number $T$ of targets and $C$ of calibrators need to be furnished. Calibrators present more steady fluxes while targets have higher flux variations in the case of rotational light curves, stellar occultations or mutual phenomena observations. Thus, PPT sets the $T$ objects with the highest flux variations as targets. The remaining $C$ objects with highest S/N values are set as calibrators. When the targets are explicitly furnished in mode 3, the remaining $C$ detected objects with highest S/N values are set as calibrators. From this point on, photometry in mode 3 proceeds automatically as in mode 2.

*3.6. Tracking fixed and moving objects*

There is no restriction about the guide, calibrators or targets being moving objects (Solar System bodies of any kind) in the FOV. PPT tracks fixed or moving objects relative to the guide (which can in turn be itself a moving object). As the object-guide relative distances $(dx, dy)$ may change with time from image to image, they can be fitted by $(dx, dy)$ polynomials in time of degree $N$ ($N$ defined by us). The relative position of each object with respect to the guide is projected to the mid-instant of the next image to be processed, so that the $(x, y)$ location in the FOV is predicted and each object can be correctly identified. We can setup parameters for the elimination of outliers in the $(dx, dy)$ polynomial fittings by a sigma-clip procedure. Fixed object-guide relative distances can be updated depending on setup. Fig. 2 brings an example of the tracking of a moving object by PPT.

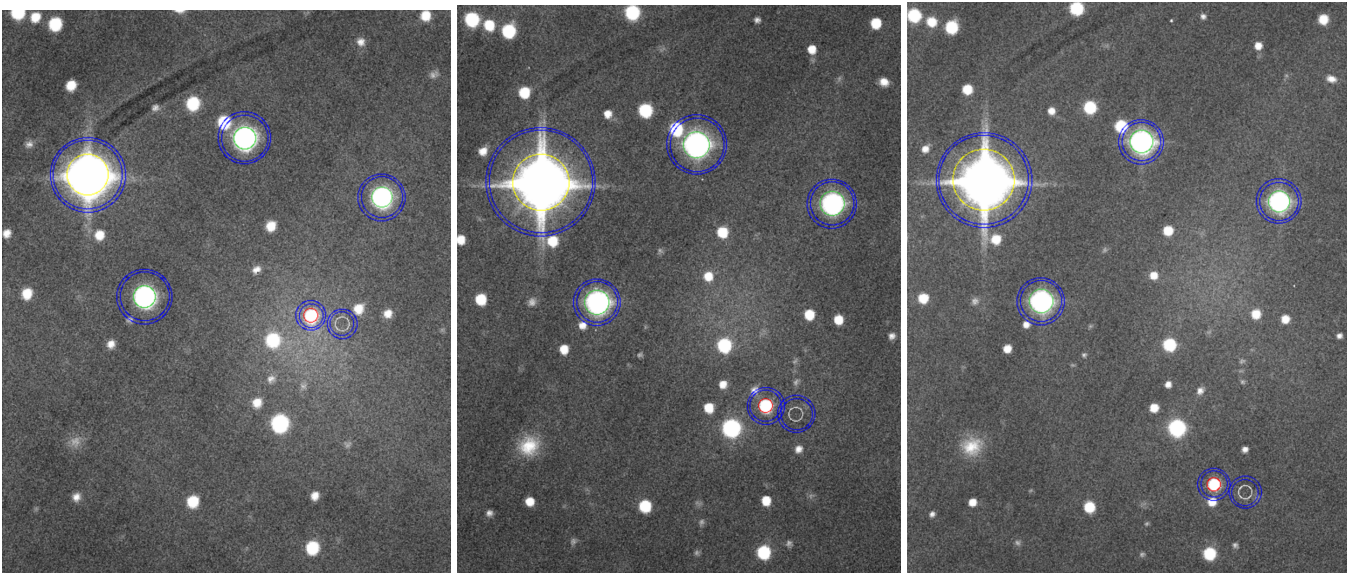

Figure 2: PPT tracking an asteroid. Trojan asteroid Priamus (`BOIA` aperture in red) is moving on a cut FOV of about 4.3′x5.9′ (east up, north left) observed on 08 July 2022 with the 0.6 m B&C telescope of the Pico dos Dias Observatory, Brazil - IAU code 874. The ghost aperture (white) follows the target, keeping a fixed distance to it. The guide (yellow) is the brightest object in the FOV. The green-aperture objects are calibrator stars keeping a fixed distance to the guide. As time goes by (from left to right panels), the computed relative positions of the asteroid with respect to the guide feed polynomial fittings in time on each $x$ and $y$ directions. The polynomials furnish the precise $(x, y)$ location of the moving object for the upcoming frames to be processed. For fixed objects (such as the calibrators in this example), the values for the relative distances can be imposed (fixed values for all images), updated from frame to frame to cope with atmospheric turbulence, bad guiding, etc, or for the same reason they can be improved frame after frame by averaging the past-computed distances.

## 4. Building light curves with PRAIA; intruding flux

A light curve (LC) from PPT presents target/calibration flux ratios as a function of time. The task produces LCs from previous PPT photometry (modes 0 – 3), or from external aperture/PSF photometry from other packages. Pertinent photometric information (target and calibration apertures, fluxes, S/N and errors, FWHM, etc) are also furnished.

During the LC building (task mode 4), PPT can automatically select the best calibrators, smooth calibration flux, and flat/normalize the flux of calibrators, targets and target/calibration flux ratios. LC smoothing and binning is also possible. Elimination filters for the LC point outliers are applied throughout all these processes, and also in the FWHM and flux errors.

PPT (task mode 5) can also compute and isolate the flux $\phi_o$ of an intruding object mixed with the object of interest in the target aperture. There are many applications for this, e.g. in the stellar occultation of bodies with atmosphere.

Fig. 3 displays a complete flowchart summarizing the photometry and LC building procedures of PPT. A step by step description of the LC building and outputs by PPT are given in the User Guide. The three available PPT methods and related outputs for computing the intruding flux $\phi_o$ are also described in detail in the User Guide. Some examples of PPT usage are shown in Section 6.

In summary, PPT outputs some LC statistics related to the selection of calibrators, calibration flux smoothing and flattening/normalization procedures performed for calibration and target fluxes, and flux ratios. LC flux ratios and errors, target and calibration fluxes, S/N and errors, LC point elimination filters, apertures, etc, are output in 3 LC files: the complete data set (main) LC, a short LC version and a user-defined LC. LC and log files are also produced in the computation of the intruding body flux $\phi_o$. PPT also outputs region files in ds9[5] format for each image for displaying the aperture and sky background rings of the guide, calibrators and targets from the photometry. Many other useful PPT features are described in the User Guide.

## 5. Digital coronagraphy with PRAIA

PDC outputs digitally coronagraphed images, and also images with the computed Source profiles superimposed with the original frames. Multi-processing is possible, i.e. many images are simultaneously coronagraphed gaining computer time.

After sky flattening, `BOIA` is used to derive precise geometric properties for the Reference Object and the Source. Precise $(x, y)$ centres are also computed with the many available PSF models, including the new PGM method.

The geometric properties of either the Reference Object or the Source furnish the shape parameters of the coronagraphy ellipse. The $(x, y)$ centres are used to set the exact location of the Source centre, i.e. the coronagraphy centre. Thin concentric ellipse rings around this centre are formed for each pixel in the FOV to be coronagraphed. Refined quartile statistics are used to compute the count that best represents each ellipse ring, and the count is then associated to each pixel. At the end, a clean Source profile is obtained, not contaminated by the nearby target nor by other objects or artifacts (diffraction spikes, etc).

---

[5]ds9 (SAOImageDS9): http://ds9.si.edu/site/Home.html

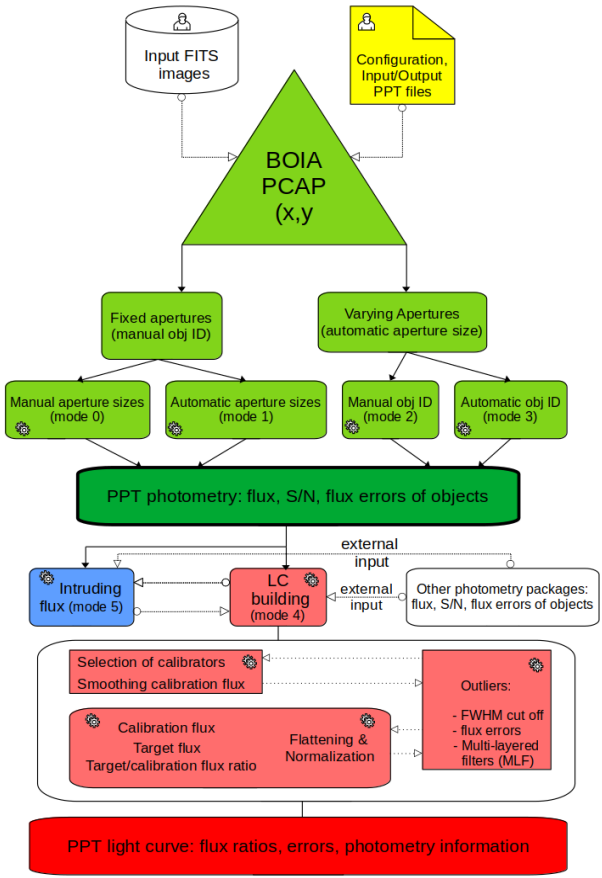

Figure 3: PPT complete workflow diagram with the photometry (green processings), light curve building (red) and intruding flux (blue) components. White processings are external in nature, configuration ones are in yellow. Person symbols represent human intervention, gear symbols indicate autonomous computer processings.

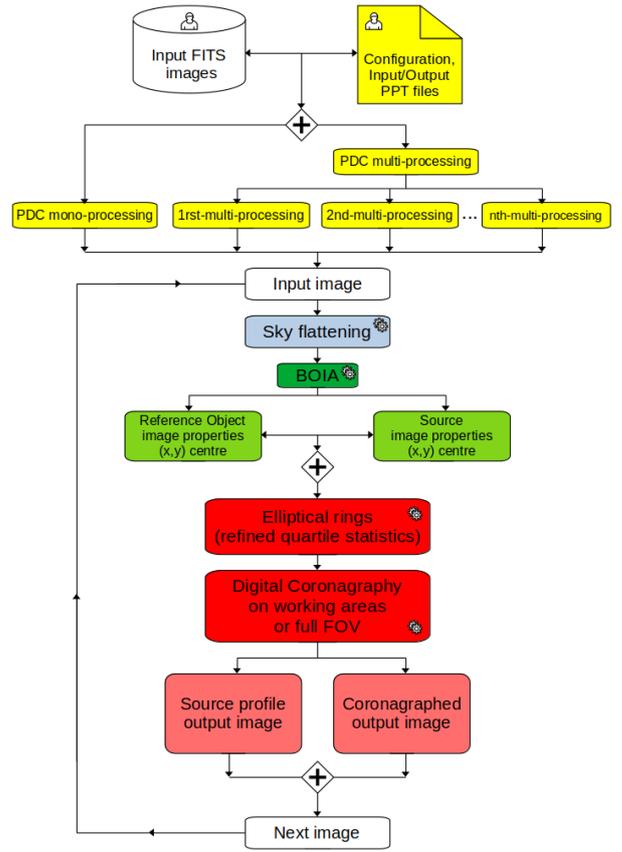

Figure 4: PDC complete workflow detailing in different colors the main components of the digital coronagraphy of one image. White processings are external in nature, configuration ones are in yellow. Person symbols represent human intervention, gear symbols indicate autonomous computer processings.

Coronagraphed images are obtained after subtraction of the original image by the clean Source profile. Coronagraphy can be done over restricted regions around the Source, over other specific regions and even for the entire FOV.

Fig. 4 displays a complete flowchart summarizing the digital coronagraphy with PDC. A detailed description of the procedures are given in Appendix A. Some examples of PDC usage are shown in Section 7.

## 6. Validation of the PRAIA photometry task

### 6.1. Solar System work done with the PRAIA photometry task

PPT is being used by the Rio Group and the Lucky Star collaboration since 2006 in the photometry of rotational light curves of asteroids and TNOs, mutual phenomena of binary asteroids and planetary natural satellites, and stellar occultations by Centaurs and TNOs. The reader will find practical usage and certification of PPT in all referred works listed in Table 1, published between 2006 – 2022. It gives the task version used at the epoch of observations (see Section 2.1), the type of event, the Solar System body class and name, the journal and the reference of the publication with the year. Most papers refer to the Rio Group and the Lucky Star international collaboration. Sometimes the use of the PRAIA package is not explicitly stated in the text, and sometimes old and quite generic references (Assafin, 2006; Assafin et al., 2011) are given.

### 6.2. Recent improvements of the PRAIA photometry task

An example of the effectiveness of the Pixelized Circular Aperture Photometry (PCAP) implemented in the last PPT version is given by the plot of two LCs from a stellar occultation by Triton observed in 05 October 2017 with the 2 m Liverpool telescope (Canary Island, La Palma, Spain)[6] in 05 October 2017 (Marques Oliveira et al., 2022).

The photometry was made with 1 calibrator star in varying aperture mode. PPT takes only about 16.5 min to run the 3000 images – 0.6s exposures of 513 x 513 pixels – with a Intel® Core™ i7-6700HQ CPU at 2.60GHz. Fig. 5 displays

---

[6] 2 m Liverpool telescope: https://telescope.livjm.ac.uk/

Table 1: 31 referred works (2006–2022) that used PRAIA photometry task.

| V | E | C | Body | Journal & Reference |
|---|---|---|------|---------------------|
| 0 | O | S | Charon | Nature: Sicardy et al. (2006) |
| 0 | M | A | (90) Antiope | Icarus: Descamps et al. (2007) |
| 1 | M | S | Uranus satellites | AJ: Assafin et al. (2009) |
| 1 | R | A | (121) Hermione | Icarus: Descamps et al. (2009) |
| 1 | O | S | Charon | AJ: Sicardy et al. (2011) |
| 1 | R | A | (216) Kleopatra | Icarus: Descamps et al. (2011) |
| 1 | O | D | Eris | Nature: Sicardy et al. (2011) |
| 1 | R | A | Binary asteroids | Icarus: Marchis et al. (2012) |
| 1 | O | D | Makemake | Nature: Ortiz et al. (2012) |
| 2 | M | S | Jupiter satellites | MNRAS: Dias-Oliveira et al. (2013) |
| 2 | O | K | Quaoar | ApJ: Braga-Ribas et al. (2013) |
| 2 | O | D | Pluto | A&A: Boissel et al. (2014) |
| 2 | O | C | Chariklo | Nature: Braga-Ribas et al. (2014) |
| 2 | O | D | Ceres | MNRAS: Gomes-Júnior et al. (2015) |
| 2 | O | D | Pluto | ApJ: Dias-Oliveira et al. (2015) |
| 2 | O | D | Pluto | ApJL: Sicardy et al. (2016) |
| 2 | O | K | 2007 $UK_{126}$ | AJ: Benedetti-Rossi et al. (2016) |
| 2 | O | K | 2003 $AZ_{84}$ | AJ: Dias-Oliveira et al. (2017) |
| 2 | O | C | Chariklo | AJ: Bérard et al. (2017) |
| 2 | O | C | Chariklo | AJ: Leiva et al. (2017) |
| 3 | O | S | Europa (J II) | A&AL Morgado et al. (2019) |
| 3 | O | K | 2003 $VS_2$ | AJ: Benedetti-Rossi et al. (2019) |
| 3 | M | S | Jupiter satellites | P&SS: Morgado et al. (2019) |
| 3 | OR | S | Phoebe (S IX) | MNRAS: Gomes-Júnior et al. (2020) |
| 3 | O | K | Varda | A&A: Souami et al. (2020) |
| 3 | O | KC | TNOs, Centaurs | A&A: Rommel et al. (2020) |
| 3 | M | T | (617) Patroclus | Icarus: Berthier et al. (2020) |
| 3 | O | C | Chariklo | A&A: Morgado et al. (2021) |
| 3 | O | S | Triton | A&A: Marques Oliveira et al. (2022) |
| 3 | O | S | Jupiter satellites* | AJ: Morgado et al. (2022) |
| 3 | O | K | Huya | A&A: Santos-Sanz et al. (2022) |

*Note*: Task version (V): 0, 1, 2 or 3 in evolution order (see Section 2.1). Event type (E): M = mutual phenomena, O = stellar occultation, and R = rotational light curve. Body class (C): D = dwarf planet, S = satellite, A = asteroid, T = Trojan, K = Kuiper Belt Object (TNO), and C = Centaur. (*): Galilean satellites Io, Europa and Ganymede.

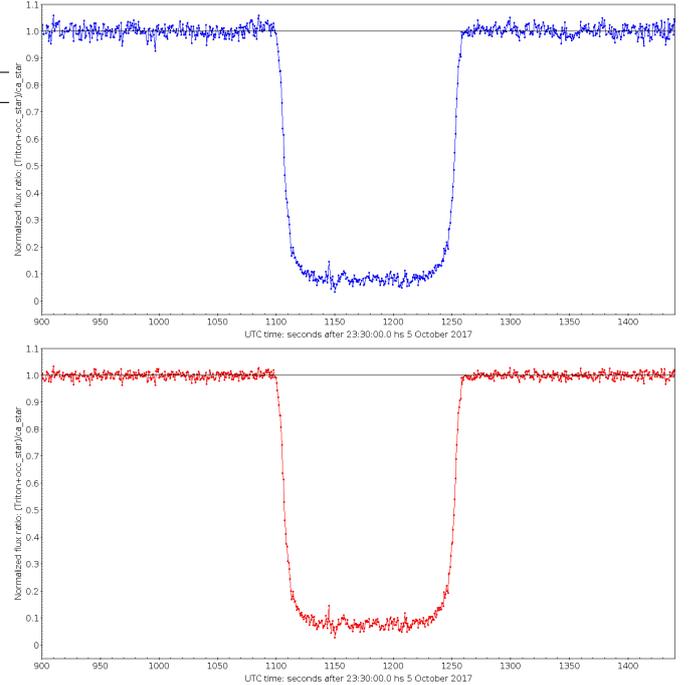

Figure 5: LCs from a stellar occultation by Triton observed in 05 October 2017 with the 2 m Liverpool telescope, generated with the previous (top panel) and current (lower panel) versions of the PRAIA photometry task (see text). Dispersion improved from 1.73 to 1.02 per cent. Credit: observations with the Liverpool telescope (Canary Island, La Palma, Spain) were made by J. L. Ortiz.

LCs obtained with the previous and current PPT versions. The dispersion (sigma) of the LC with version 2 was 1.73 per cent. The same data and parameters were used to obtain the LC using the last PPT version 3, with a dispersion (sigma) of 1.02 per cent – an improvement of 41 per cent over the previous LC.

In our tests, we usually improved the error of the LCs (dispersion measured by the sigma of the LC) by 30 – 60 per cent with respect to the previous task version.

### 6.3. Intruding body flux in a stellar occultation: an example

As an example for the computation of the intruding body flux $\phi_o$ (Section 4), we again refer to the stellar occultation by Triton in 05 October 2017 (Marques Oliveira et al., 2022) observed with the Liverpool telescope.

Two blocks of intruding flux calibration images where acquired in the night before the event with the occultation star separated from Triton in the FOV. Observations were made near the same airmass of the occultation itself. The first block (301) was observed with the CCD inverted in $Y$ with respect to the orientation angle of the occultation observations, while a few minutes later block 302 was observed with the same CCD orientation as that of the event, and at a closer air mass. Fig. 6 displays the FOV of occultation and calibration observations.

Fig. 7 displays the normalized occultation LC at the occultation instant $t_0$, the FLC (Fraction Light Curve) with isolated body flux and other 3 auxiliary LCs generated for each calibration block for the computation of Triton's intruding body flux $\phi_o$ from Liverpool telescope observations (see Section 4 and details in the User Guide). The FLC obtained from the separated fluxes of Triton and occultation star with calibration images serves as a sanity check, because they should closely match the light curve flux ratio of the event. From Fig. 7, the FLC of block 302 passes the sanity check, but that of block 301 does not. The CCD observations of block 302 were made with the same position angle as that of the occultation, while the block 301 orientation was different. Also, the air mass of block 302 is closer to that of the occultation than that of block 301. Therefore, the computation of $\phi_o$ was made by using the calibration images of block 302 only.

For example, let's compute $\phi_o$ following Method 1 (see Section 4 and details in the User Guide). The relative error of the normalized LC flux ratio $f$ for the isolated occultation star from block 302 was 1.45 per cent. The same air mass (height=51.9$^o$) as that of the occultation is achieved at $t_c = 22^h\ 28^m\ 30^s\ 04$ October. From the polynomial fittings in flux $f$ for this instant, $f_c = 0.4391 \pm 0.006$. The mid-instant $t_0$ of the occultation is $t_0 = 23^h\ 49^m\ 37^s$ 05 October. From the polynomial fittings in the normalized flux ratio $F$ of the occultation light curve itself, we got a relative error of 1.02 per cent. For $t_0$, we have $F_0 = 0.6614 \pm 0.007$. Using the relations described in Method 1, we compute $\phi_o = 0.336 \pm 0.015$. Retrieving the clean flux of the occultation star with $\phi_o$ allows the use of PPT for obtaining the entire occultation LC without the intruding flux of Triton (see Section 4 and details in the User Guide). Fig. 8 compares the occultation LCs with and without the intruding flux for the Liverpool telescope. The occultation star flux is non zero during the Triton occultation due to refraction by its atmosphere. In fact, measuring this residual flux is of uttermost importance to model the body's atmosphere.

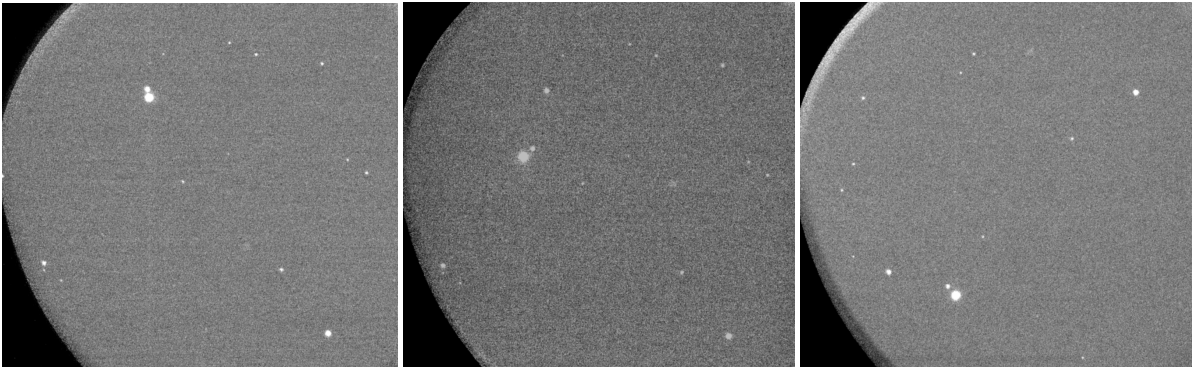

Figure 6: FOV of Liverpool telescope observations from the stellar occultation by Triton in 05 October 2017 (left). Block 302 (middle) and 301 (right) intruding flux calibration images were taken at the night before, near the same air mass as that of the occultation. The CCD orientation of the FOV is the same for occultation and block 302 observations and is inverted in $Y$ with respect to block 301 images. Neptune and Triton are at the top-left corner in the FOV of the occultation, already obstructing the occultation star which is visible above them in the FOV of block 302; the calibrator star appears at the lower-right corner of these FOVs. In the FOV of block 301, Neptune/Triton and the occultation star appear at the lower-left corner and the calibrator star at the upper-right one. Credit: observations with the Liverpool telescope (Canary Island, La Palma, Spain) were made by J. L. Ortiz.

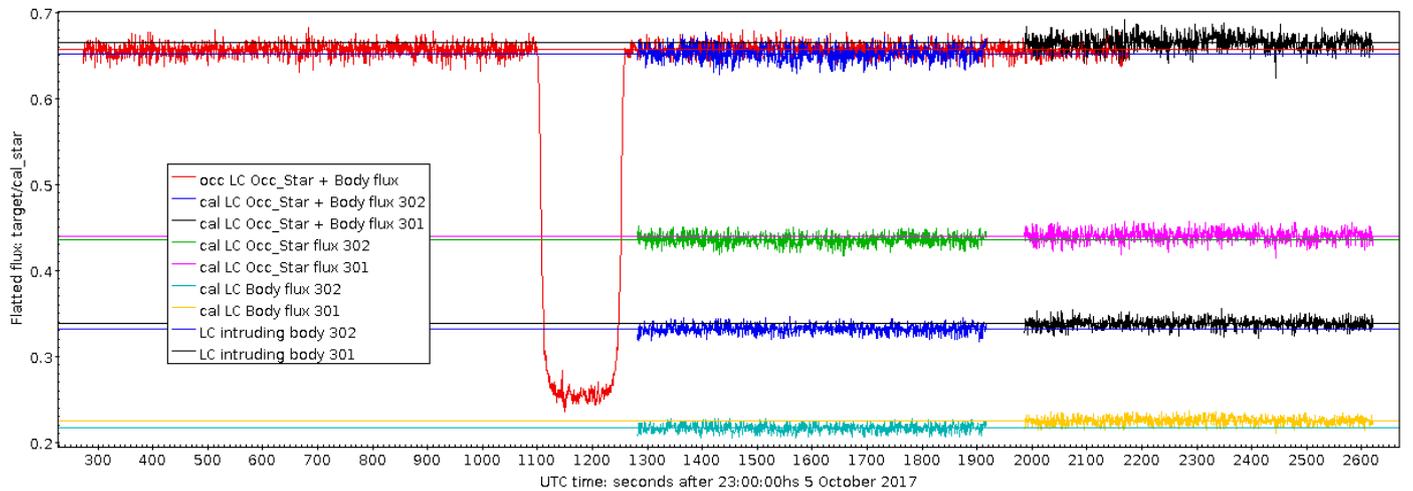

Figure 7: Normalized occultation LC at $t_0$, FLC (Fraction Light Curve) and other 3 auxiliary LCs from calibration and occultation observations with the Liverpool telescope (see Section ??). Two blocks of calibration images are treated separately: blocks 301 and 302 (see text). The instants of calibration block observations were converted to instants nearby the occultation, so as to exactly match the air mass of each calibration observation with the corresponding air mass of the occultation observations at the plotted instants. The air mass of block 302 observations is closer than that of block 301 with respect to the air mass of the occultation window. Credits: observations with the Liverpool telescope (Canary Island, La Palma, Spain) were made by J. L. Ortiz.

6.4. *PRAIA x DAOPHOT comparisons*

The DAOPHOT package (Stetson, 1987) under IRAF (Tody, 1986, 1993) is widely used in crowded fields, when PSF photometry is recommended. The moderate crowded field in Fig. 9 allows for a nice PPT x DAOPHOT photometry comparison.

Varying aperture photometry with 16 calibrators using PPT and PSF photometry using DAOPHOT/IRAF with 11 calibrators result in the LCs shown in Fig. 10. The target and comparison star LCs match exactly after respectively applying zero-point offsets of -0.68 and +2.45 for the corresponding DAOPHOT/IRAF LCs. The dispersion of the comparison star LC is 0.130 for PPT and 0.132 for DAOPHOT/IRAF. Both tasks render the same results regardless of the different calibrators, internal procedures and photometry used – aperture versus PSF measurements.

While by construction PPT can easily deal with moving objects, from moderate (see Fig. 2) to fast apparent speeds (e.g. Near Earth Asteroids), DAOPHOT/IRAF is designed for fixed objects. However, Spanish researchers within the Lucky Star collaboration managed to develop IDL (Interactive Data Language) routines that perform photometry with `DAOPHOT` for FOV-slow-moving objects such as TNOs (Ortiz et al., 2020; Fernández-Valenzuela et al., 2016), allowing for a direct comparison with PPT. Usually, similar results are obtained. Among some examples present in the works listed in Table 1, we can cite the most recent one (Santos-Sanz et al., 2022), where both pipelines were used in the photometry of stellar occultation and rotational light curve observations.

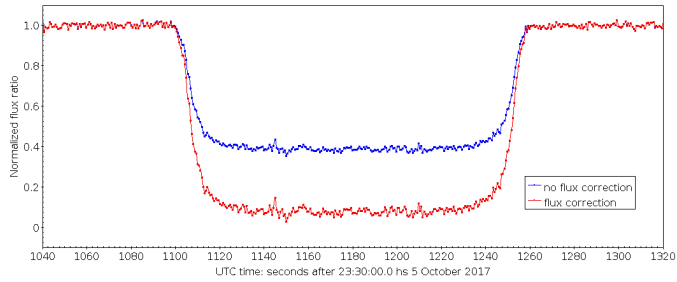

Figure 8: Liverpool normalized LCs with and without contamination by Triton's flux. As expected, in the LC without contamination some occultation star flux is still present (non zero) during the occultation, due to refraction by Triton's atmosphere. Credits: observations with the Liverpool telescope (Canary Island, La Palma, Spain) made by J. L. Ortiz.

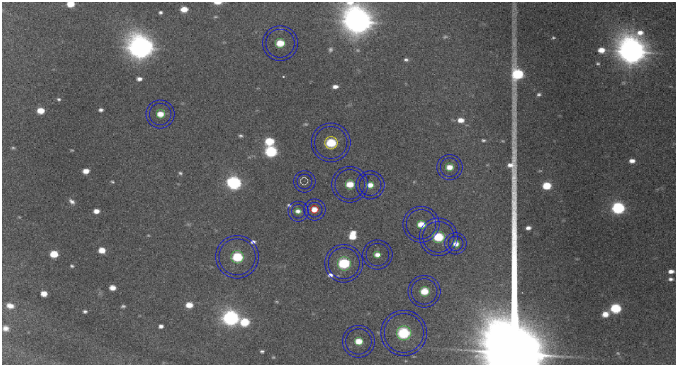

Figure 9: FOV centred at the Cataclysmic Variable star MLS2043 of the Mount Lemmon Survey, part of the Catalina Real-Time Transient Survey (Drake et al., 2009, 2014). The CCD frames were taken with the 0.6 m B&C telescope of the Pico dos Dias Observatory, Brazil – IAU code 874. Circled objects are 16 calibration stars. The target is indicated by the red circle and the comparison star is the circled star to its left, with the ghost right above it (white circle). Blue circles indicate sky background rings. Credits: Observations by C. V. Rodrigues and A. S. de Oliveira.

In cases with very faint or blended targets in the images, PPT outperforms `DAOPHOT/IRAF`. A critical example is described in Fig. 11 for the stellar occultation on 5 June, 2019 by the TNO Quaoar. This event was used in the discovery of a ring outside the Roche limit around Quaoar (Morgado et al., 2023). The target (body + occultation star) was very faint – V magnitudes 17.4 and 18.8 for star and Quaoar, respectively. The occultation star was also blended with a field star, as shown in the FOV displayed in Fig. 11. In this case, `DAOPHOT/IRAF` cannot deliver a light curve – it crashes, unable to fit a PSF to the blended, faint target. On the other hand, PPT delivers a light curve capable of detecting the faint QR1 ring within a $5\sigma$ confidence level, as shown in Fig. 11.

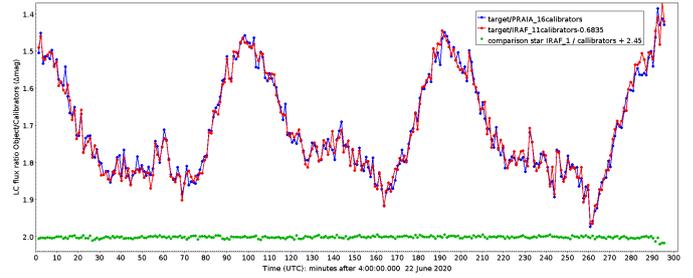

Figure 10: Target LCs from varying aperture photometry with 16 calibrators using PPT (blue), and from PSF photometry using `DAOPHOT/IRAF` with 11 calibrators minus 0.68 (red), for the observations of the FOV shown in Fig. 9. The respective LCs of the comparison star are also indistinguishable from each other after adding 2.45 for the `DAOPHOT/IRAF` one.

*6.5. PRAIA x other popular Solar System photometric packages*

Two differential aperture photometric programs are popularly used at PPT's niche to the best of our knowledge, i.e. in stellar occultations and rotational light curve observations: Tangra[7] (Pavlov, 2020) and Pymove[8] (Anderson, 2019).

Tangra is specially good in the manipulation of video-camera observations. Pymove also performs astrometry. Although these packages deserve merit, the implemented photometry algorithms and procedures are not yet clearly described in their documentation. And there are no published referred articles certifying Tangra or Pymove yet. Tests performed with these packages along our collaboration work with professionals and the amateur community in some stellar occultation campaigns showed that PPT gives systematically better LCs. Above all, these packages are designed for human intervention, making it very difficult to deal with great amounts of observations, not to mention heterogeneous ones.

Although by purpose not appealing from a Graphical User Interface (GUI) aesthetic point of view, PPT may become a very attractive photometric tool for the amateur community in the future because of the simplicity in the usage and fast performance, even for great amounts of heterogeneous observations. Noting that it is a well documented photometry package certified by specialists in the field.

## 7. Validation of the PRAIA Coronagraphy task

*7.1. Examples*

PDC's digital coronagraphy is insensitive to diffraction spikes, bright or faint objects nearby the Source (like satellites or stars), saturation artifacts in the Source profile and even to the existence of anomalous darker or brighter regions. This is because higher or anomalous lower count pixels are eliminated in the quartile statistics of the rings and thus do not affect the computation of the light profile of the Source. In fact, most of these image artifacts are revealed, appearing nearby or at the inner parts of the Source in the coronagraphed images after the removal of the light profile of the Source.

---

[7]Tangra: www.hristopavlov.net/Tangra/Photometry.html
[8]Pymove, under the auspices of IOTA, International Occultation Timing Association: https://occultations.org/sw/pymovie/PyMovie-doc.pdf

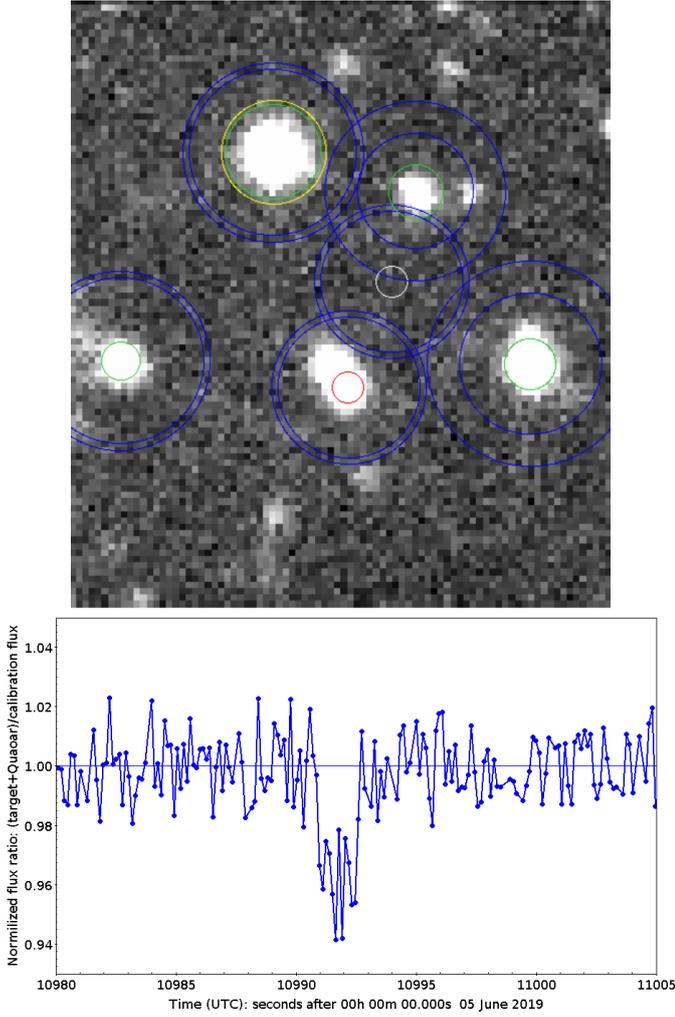

Figure 11: PPT LC sample from the newly discovered QR1 ring of TNO Quaoar (Morgado et al., 2023) obtained with the stellar occultation on June 5, 2019 observed in filter I (0.13 s exposures) with the Hipercam of the Gran Telescopio Canarias of 10.4m (GTC, Canary Islands, La Palma, Spain). The top panel shows a FOV of about 20″ x 25″ (north up, east right) with the target (V = 17.4 star + V = 18.8 Quaoar, red aperture) blended with a field star, yet separated enough for PPT photometry without the need of PDC digital coronagraphy. Yellow apertures indicate the guide, green calibrators, and white is the ghost aperture. Blue circles indicate sky background rings. The LC in the lower panel shows the starlight dip due to the QR1 ring, after the main body's occultation not shown in the plot. LC dispersion (1$\sigma$) was 0.93 per cent. The extremely tenuous 5 per cent light dip of the QR1 ring could be nevertheless detected by PPT with a 5$\sigma$ confidence level. Credits: GTC observations by J. L. Ortiz.

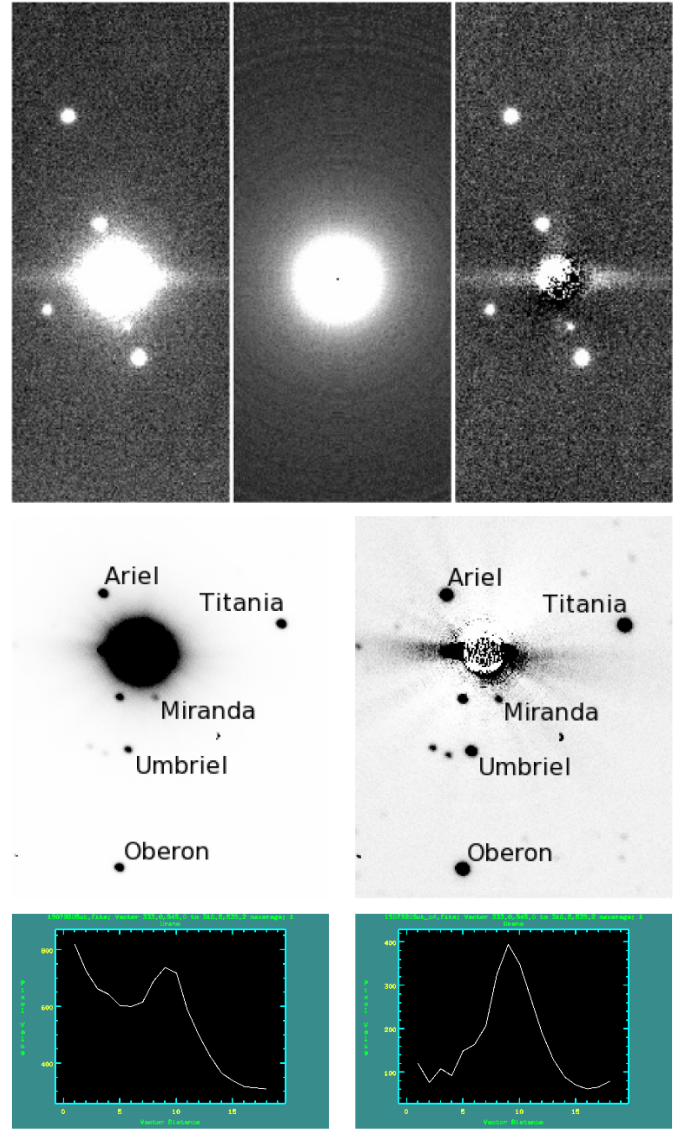

Figure 12: Digital coronagraphy of Uranus with its satellites observed with the 1.6 m telescope at Observatório do Pico dos Dias – OPD/LNA, Brazil (IAU code 874). Top panel: from left to middle to right we have the original image, the obtained Uranus brightness profile and the output coronagraphed image. The satellites aligned from top to bottom are Titania, Ariel, Miranda, Umbriel and Oberon (the last two are indistinguishable from each other). Middle panel: we have for another OPD observation the inverted color map images before (left) and after (right) the digital coronagraphy. Lower panel: corresponding light profiles of Miranda along a segment linking Uranus and Miranda centres before (left) and after coronagraphy. Credits: top panel from Assafin et al. (2009), middle and lower panels from Camargo et al. (2015).

Typical examples of the application of the digital coronagraphy by PDC are illustrated in Fig. 12 (Uranus system) and Figs. 13 and 14 (Neptune system).

Fig. 12 indicates how digital coronagraphy can improve astrometric measurements. Before it, it was almost impossible to correctly delimit Miranda under the scattered light of Uranus. The $(x, y)$ measurements of contaminated light profile are strongly dependent on the modeling of the high-gradient counts of the scattered light surrounding the satellite. After the coronagraphy, we clearly isolate Miranda from the brighter Source (Uranus) and successfully retrieve the light profile of the satellite. Consequently, the $(x, y)$ measurements are improved.

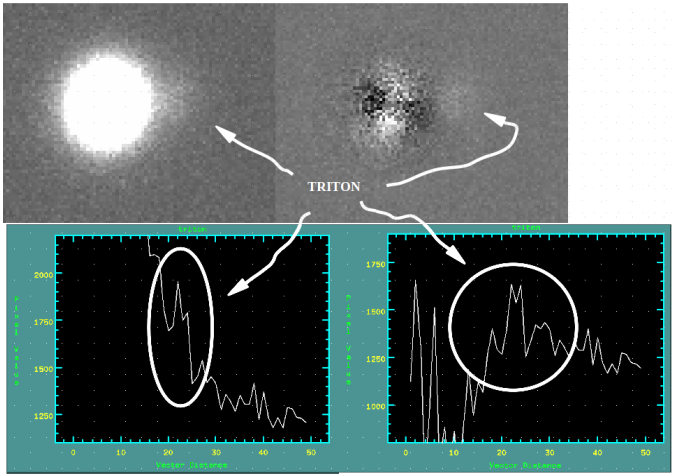

Figure 13: Top left panel shows Neptune with Triton barely visible at its right. Top right panel brings the coronagraphed image with Triton now more clearly distinct in the image. A f/10 telescope of D=81 cm was used. The sky conditions were very poor (seeing about 5" and strong winds). Below, we see the corresponding Triton profiles at the same segment linking Neptune and Triton centres. After the coronagraphy, the PSF of Triton is now more clear and symmetric above the flatted sky background. Credits: the original images were furnished by B. Sicardy, see Marques Oliveira et al. (2022).

Fig. 13 illustrates how digital coronagraphy can save photometric observations under poor sky conditions. It was impossible to do the photometry of Triton due to the small separation to Neptune, from a set of short-focus telescope observations under strong seeing. The situation is dramatically improved after digital coronagraphy. Fig. 14 illustrate light curves obtained before and after digital coronagraphy for a stellar occultation by Triton using the same set of observations from the FOV displayed in Fig. 13. We go from a non-event detection, high-noise light curve situation to an event detection, low-noise light curve.

The task works fine for well-tracked round images or moderately elongated ones. But even if it is not the case, a good removal of the scattered light around the Source is still obtained, evidencing the power of elliptical rings in PDC's digital coronagraphy. Fig. 15 displays such example for Uranus. The planet's and satellites' profiles are clearly elongated due to bad guiding. Even so, the coronagraphy successfully suppressed the scattered light, enhancing Miranda's profile.

### 7.2. Solar System work done with the task

PDC is used by the Rio Group since the 90's and under the Lucky Star international collaboration since the 2010's in the astrometry and photometry of natural satellites and stellar occultations by dwarf planets and TNOs. Practical usage/certification of PDC is found in all referred works listed in Table 2. It gives the task version used at the epoch of observations (see Section 2.2), the study type, the Solar System body class and name studied, the journal and the reference of the publication with the year. In the case of satellites, the host planet was coronagraphed. In the case of Pluto, the occulted star was for calibration purposes. Sometimes the use of the PRAIA package is not explicitly stated in the text, and sometimes old and quite generic references (Assafin, 2006; Assafin et al., 2011) are given.

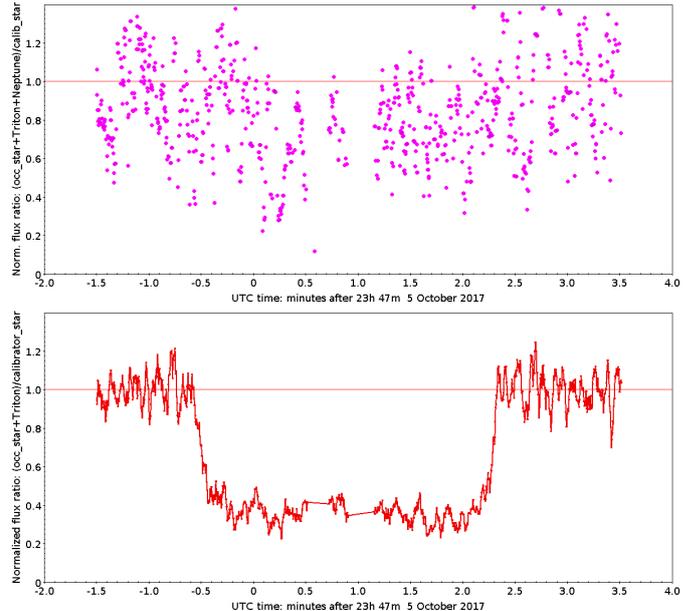

Figure 14: Top panel. Tentative light curve with the original images (no coronagraphy, see Fig. 13). The target flux is severely contaminated by Neptune's light. No stellar occultation flux drop can be detected. Lower panel. Light curve with coronagraphed images (Fig. 13). The flux drop of a stellar occultation is clearly seen. The normalized flux ratio error was 8.8 percent. Photometry was made with the PRAIA photometry task. Credits: raw images furnished by B. Sicardy; see Marques Oliveira et al. (2022).

### 7.3. Other digital coronagraphy tools and methods

Few digital coronagraphy algorithms were published: the Angular Differential Imaging (ADI, Marois et al. 2006), the Locally Optimized Combination of Images (LOCI, Lafrenière et al. 2007), the Karhunen-Loève Image Projection (KLIP, Soummer et al. 2012) and the Digital Coronagraph Algorithm (DCA, Valle et al. 2018). All these algorithms rely on PSF subtraction. For all of them, deep optical knowledge of high quality telescopes is necessary for PSF reconstruction, with the need to build external specific reference image data sets per run. ADI is only applicable for altitude/azimuth telescope observations with the field derotator switched off. DCA mimics the interference of a coronagraph instrument, being designed for space telescopes at the diffraction limit for which the PSF is well known a priory. Unfortunately, none of these algorithms can in practice be used in routine works with voluminous observations from heterogeneous instruments under variable sky conditions.

Worth mentioning is the Larson-Sekanina (LS) digital coronagraphy procedure first applied for digitized images of comet Halley (Larson & Sekanina, 1984) and latter developed and applied for other comets. The method basically consists of producing auxiliary $A$ images by radial and angular pixel shifts followed by subtraction of the original $O$ image by the $A$ ones to get coronagraphed $C$ images. By construction, LS is suited for enhancing sparse large-scale faint structures around the coronagraphed Source, and that is why it is used in the study of comet

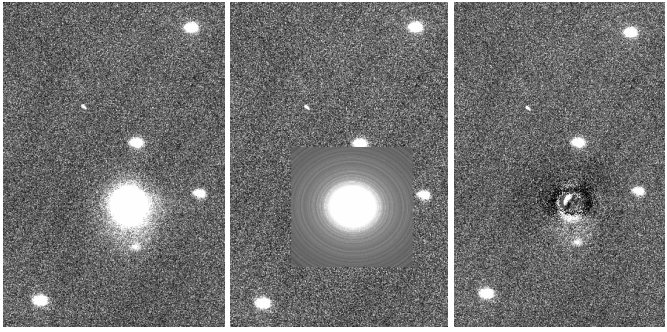

Figure 15: Uranus and satellites observed with a methane filter at the 1.6 m OPD telescope. Miranda is the faint object below the planet. Planet and satellite profiles are elongated due to bad guiding. Left: original image. Middle: Uranus' profile obtained by digital coronagraphy. Right: coronagraphed image. Even under severe image elongation, the scattered light is successfully removed showing the power of elliptical rings in PDC's digital coronagraphy.

Table 2: 8 works (1995–2022) that used PRAIA digital coronagraphy task.

| V | T | C | Body | Journal & Reference |
|---|---|---|------|---------------------|
| 0 | A | S | US | A&A: Veiga & Vieira Martins (1995) |
| 1 | M | S | US | AJ: Assafin et al. (2009) |
| 2 | A | S | US | A&A: Camargo et al. (2015) |
| 2 | O | D | Pluto | ApJ: Dias-Oliveira et al. (2015) |
| 3 | M | S | Amalthea | P&SS: Morgado et al. (2019) |
| 3 | A | S | US | MNRAS: Santos-Filho et al. (2019) |
| 3 | O | S | Triton | A&A: Marques Oliveira et al. (2022) |
| 3 | A | S | US | P&SS: Camargo et al. (2022) |

*Note*: Task version (V): 0, 1, 2 or 3 in evolution order (see Section 2.2). Study type (T): M = mutual phenomena, O = stellar occultation and A = astrometry. Body class (C): D = dwarf planet, S = satellite. US = Uranus main satellites.

observations. However, LS is not suited for enhancing the faint profile of small-scale round-shaped point-like objects contaminated by a close-by bright profile Source. Sub-pixel shifts should be necessary, introducing the problem of extrapolation with no guarantee of good results. Also, it is hard to find the best radial-angular shift for a given small profile, following an automated process. Above all, the resulting profile is deformed in the process, rendering it useless for high precision astrometric or photometric measurements. The situation is even worse when it is the faint object profile that we want to coronagraph to get a cleaner profile for the brighter nearby object.

PDC overcomes all the problems here reported with its build-in capacity of deriving coronagraphed images for observations taken from any type of instrument without any a priory knowledge or data in a fast, robust and efficient way, even under severe seeing or bad guiding. PDC successfully enhances the profiles of objects only a few arcseconds away from the coronagraphed Source, propitiating precise astrometric and photometric measurements of the observed targets.

## 8. Conclusions, remarks, future implementations

We presented the photometry and digital coronagraphy concepts under `PRAIA` – astrometry is detailed in another paper (Assafin, 2023). `PRAIA` gives photometry/astrometry support for the studies of Solar System bodies carried out by the Rio Group and the international collaboration under the Lucky Star Project. The standalone codes have no dependencies with external packages and are written in `FORTRAN`. They are simple to install and use. The package does not suffer from deprecation issues as is common with Python-based packages. Unlike packages such as `IRAF` which lost support, `PRAIA` has constant updates with new releases and, from time to time, upgrades to new versions since 1992 (astrometry) and 2006 (photometry).

The `PRAIA` photometry task (PPT) and the `PRAIA` digital coronagraphy task (PDC) perform accurate/precise/fast differential aperture photometry and digital coronagraphy with full automation over huge amounts of observations from heterogeneous instruments. PPT is suited for the photometry of rotational light curve, mutual phenomena and stellar occultation observations, and PDC support the astrometry and photometry of natural satellites, mutual phenomena and stellar occultations.

PPT was used in 31 publications between 2006 – 2022 (Table 1). It brings innovations such as: a) Pixelized Circular Aperture Photometry (PCAP) – a fast robust and precise aperture photometry method that avoids flux computations by pixel sub-sampling or fractioning; b) `BOIA` (Browsing Objects: Identification and Analysis) which automatically identifies objects and the sky background without using a priory information or sigma factors, and optimizes aperture sizes avoiding subjective FWHM factors; c) Photogravity Center Method (PGM) – a new centring method better than Modified Moment (Stone, 1989).

PDC was used in 8 publications between 1995 – 2022 (Table 2). Latest task improvements include: a) coronagraphy of faint objects close to bright ones and vice-versa with the introduction of the Reference Object and the Source; b) use of elliptical rings which allow for the coronagraphy of elongated profiles; c) refined quartile statistics, improving coronagraphy results; d) better object detection and characterization with the novel method `BOIA`; e) better astrometry improving the coronagraphy centre, including the new centring method PGM among other circular and elliptical Gaussian and Lorentzian profiles; f) multiprocessing capabilities, improving computer processing speed.

To the best of our knowledge, PPT is unique in terms of its capability to automatically do fast precise differential aperture photometry for large amounts of heterogeneous observations without any a priory knowledge or external calibration data, specially in Solar System photometry. PPT proved to be superior to other popular software used in this niche like Tangra and Pymove. PPT light curves have less dispersion, and are obtained without the typical need of human intervention of these packages, which considerably slows down the efficiency and possibly degrades the photometry. Also, comparisons show that PPT's photometric precision equals that of `DAOPHOT` – a standard in astrophysical photometric work. PPT outperforms `DAOPHOT` in the photometry of very faint and/or blended objects.

Unlike more specialized digital coronagraphy algorithms such as ADI, LOCI, KLIP and DCA, PDC is capable of deriving coronagraphed images for observations made with any type of instrument, without any a priori knowledge or external calibration data. PDC is also superior to the Larson-Sekanina method (LS) mostly used in cometary studies. PDC performs at least equally well and faster at large scales, but unlike LS it also successfully enhances the profiles of objects only a few arcseconds away from the coronagraphed source. Besides, PDC preserves the original counts and photometric properties onto the coronagraphed images in both cases, contrary to LS.

Besides Solar System work, PPT can be used in the photometry of variable and cataclysmic stars and transient phenomena like exoplanet transits and microlensing, when only relative flux is needed. PDC can also support astrophysical observations.

Future PPT updates include adding centring algorithms such as the elliptical Gaussian PSF, and general circular and elliptical Lorentzian PSFs. In turn, we also foresee the use of elliptical apertures for PPT, already implemented in PDC.

These PRAIA tasks run under MacBook and debian-based Linux operation systems. They were thoroughly tested with Ubuntu distributions 16.04 to 22.04. We plan to encapsulate the codes in docks so that they can be run on any Linux and Windows systems. We also plan to run the tasks as photometry and digital coronagraphy services to the astronomical community in the LIneA Solar System Portal[9]. In the future all PRAIA package will be available trough GitHub[10]. Currently, PPT and PDC task codes, input files and documentation are publicly available for the first time at https://ov.ufrj.br/en/PRAIA/.

**Declaration of competing interest**

The author declares not having any known competing financial interests or personal relationships that could have appeared to influence the work reported in this paper.

**Acknowledgements**

I thank Editor Maria Cristina De Sanctis and the two anonymous reviewers who made significant contributions for improving the text. I also thank Dr. J. I. B. Camargo, Dr. F. Braga Ribas, Dr. R. Vieira Martins, Dr. G. Benedetti Rossi, Dr. S. Santos Filho and Dr. C. H. Veiga for the fruitful discussions in some aspects of the tasks' upgrades, and for helping in the internal revision of the text. I acknowledge CNPq grants 427700/2018-3, 310683/2017-3 and 473002/2013-2. The Lucky Star Project agglomerates the efforts of the Paris, Granada and Rio teams, and is funded by the European Research Council under the European Community's H2020 (ERC Grant Agreement No. 669416).

---

[9]LIneA stands for Interinstitutional Laboratory for e-Astronomy. LIneA Solar System Portal: https://tno-dev.linea.org.br/
[10]GitHub: https://github.com/

**Appendix A. Digital coronagraphy with PRAIA**

*Appendix A.1. Sky background flattening*

Prior to the coronagraphy, the sky background of the whole FOV is flattened. This levels the sky background of the FOV or at least of large FOV portions around the coronagraphy working area. Strong sky background gradients may jeopardize the ring statistics, thus affecting the coronagraphy results.

The sky background is modeled by a bi-variate polynomial in $(x, y)$ of degree $N$. Assuming that the majority of pixels in the FOV represent the sky background, we initially fit the polynomial to the whole FOV. The fitted pixel counts are previously smoothed by a binomial filter of $(2n+1)$ channels (we use n = 5). The polynomial is then normalized and the original input image is flattened by dividing the original non-smoothed counts by the normalized polynomial, pixel by pixel. Then, a histogram analysis is done for determining the mode $M$ and a dispersion $S$ for the sky background – the mode is associated to the typical sky background value for the whole FOV. The polynomial fit is then repeated, taking only sky background pixels within 2.5 $S$ from $M$. The polynomial is normalized and the image is again flattened for the last time. One last histogram analysis and sky background polynomial fit is done. This polynomial represents a modeled sky background for the whole FOV. It has 2 utilities: a) mask counts for bad pixels; b) mask counts for coronagraphed pixels that eventually resulted in negative counts after the end of the coronagraphy process.

Slight vignetting can usually be dealt with by low order polynomials of degree $N=2$, and moderate vignetting by $N=4$. Frequently, no sky background flattening is truly needed, so just using $N = 0$ works. Even so, the sky background level and dispersion are determined and a flat sky background solution is computed and used to mask the counts of bad pixels or negative coronagraphy pixels.

After coronagraphy, the flattening is reverted and the original sky background restored, i.e. the original photometric properties are fully preserved onto the coronagraphed images.

*Appendix A.2. The Reference Object*

The Reference Object serves as reference for the $(x, y)$ location of the Source and for the setup of the coronagraphy ellipse. It should have a high S/N and be a well-imaged isolated object not far from the Source. It must be the brightest object in the valid FOV range. It can also be the Source itself.

PDC automatically finds and determines the $(x_R, y_R)$ centre of the Reference Object (see Section 2.3). Many high-level centring algorithms are now available: the new Photogravity Center Method (PGM), 2 Circular PSFs and 2 Elliptical ones, Gaussian or Lorentzian PSFs, fitted to the pixels selected by BOIA.

We can also directly input the $(x_R, y_R)$ centre of the Reference Object and keep it fixed for all images – this option is only useful in test cases or in very specific occasions.

We also have the option to fix the relative distance between the Reference Object and the Source for all images. In this case, the $(x_R, y_R)$ centre is input together with the Source's $(x_C, y_C)$ coordinates.

A central moment analysis is made and the ellipse's eccentricity, semi-major and semi-minor axes, and the orientation angle of the Reference Object are stored. If one of the two elliptical PSF models is chosen as the centring algorithm, the corresponding values are also stored. We can also impose these parameters, which then remain fixed for all images. These values may be used to setup the coronagraphy ellipse.

*Appendix A.3. The Source*

The Source is the object in the FOV which we want to coronagraph. It can be any object in the FOV, regardless of the brightness. It can even be the Reference Object.

The same Reference Object options (Appendix A.2) are available for imposing or determining the ($x_C$, $y_C$) Source centre, ellipse's eccentricity, semi-major and semi-minor axes, and the orientation angle. Input Source coordinates may be used with Reference Object ones to fix the relative distance for all images. This relative distance may also be used as a kick off, being continuously updated in order to estimate for each image an initial ($x_i, y_i$) input for finding optimum BOIA apertures and updating ($x_C$, $y_C$) Source centres (see Section 2.3).

*Appendix A.4. Centre and ellipse for coronagraphy*

Here, we explore the possible centre/ellipse setups of the Source and Reference Object. For the best coronagraphy result, we must get as accurately and precisely as possible the ($x_C$, $y_C$) centre of the Source and ellipse parameters that best represent the Source's profile.

PDC allows for a very flexible setup for defining the ($x_C$, $y_C$) centre of the Source that will be used as the coronagraphy centre. There are three possibilities for obtaining this centre: 1) it is imposed; 2) it is derived from Source centring; 3) it is derived from Reference Object centring only.

In case 1, imposing the ($x_C$, $y_C$) Source centre is only useful in tests or very specific cases, as it is only valid for one image.

In case 2, we first set an initial relative distance ($dx, dy$) between the Source and the Reference Object. It is first computed prior to the processing of the first image from the input ($x, y$) coordinates of the Reference Object and the Source. These coordinates may be taken from the inspection of any image – preferably from a well exposed one. The only request is that they be estimated from the same image for astrometric consistency. After improving the ($x_R$, $y_R$) centre of the Reference Object with the setup centring algorithm, the initial relative distance ($dx, dy$) is used to estimate the ($x, y$) location of the Source. Using this ($x, y$) location, the Source's chosen centring algorithm is applied to get the final improved ($x_C$, $y_C$) coronagraphy centre. The new improved ($x_R$, $y_R$) and ($x_C$, $y_C$) coordinates are used to update the relative distance ($dx, dy$). The process is repeated for all images. Relative distances are updated at each image and used in the next. Or the first updated distance is fixed and kept for the remaining images.

In case 3, we do not make any distance updates at all. No Source centring is actually made and the ($x_C$, $y_C$) coordinates are obtained from the input distance and by the centring of the Reference Object. Notice that in case 2, when we update the distance for the first image only, a Source centring is performed although only for the first image.

Choosing the most adequate case depends on the circumstances. If the Reference Object is also the Source and is a bright object – like a planet – then case 3 suits perfectly. Case 3 might also be suited for the coronagraphy of a faint star very close to a brighter target star, on which case the Reference Object should be another isolated object – not the same star as the Source or the target.

Regarding to the definition of the ellipse parameters to be used in the coronagraphy – the eccentricity and the orientation angle – the task also allows for a very flexible setup. There are two classes of solutions, one involving the Source itself, the other involving the Reference Object. On each class, there are three possibilities for setting the ellipse parameters: 1) direct input; 2) from central moment analysis; 3) from PSF fitting.

Imposing ellipse parameters through either the Reference Object or Source setup is useful for circular rings, on which case eccentricity is zero (using circular PSF models has the same effect). Otherwise, it is only useful in tests or specific cases.

Choosing ellipse parameters from moment analysis or elliptical PSF fittings generally render similar results. However, the moment analysis may encompass pixels that are excluded in the PSF fitting (hot pixels, spurious counts, etc), so that in principle PSF fittings render better results. On the other hand, depending on the nature of the object (isolated Reference Object versus Source with nearby targets), the S/N and surrounding contaminants, the moment analysis parameters may be preferable.

One must be cautious when choosing the object to extract the ellipse parameters for the coronagraphy of the Source. The most natural choice – the Source itself – may not be the best. There are cases when the brightness profile of the Source is somewhat (sometimes severely) compromised. This is the case when the Source is a fainter star partially embedded by a close brighter target. In this case, the moment analysis of the Source may be corrupted and the PSF fitting may be impossible. In such cases, the use of the ellipse parameters of the Reference Object is recommended. The underlying assumption is that the PSFs of objects not far away from each other in the FOV are the same. Thus, if the ($x_C$, $y_C$) is ok, the result is the same as with the PSF of the Source itself (if known). Moreover, as the Reference Object is usually a bright and isolated object, its PSF is better resolved and usually furnishes more precise and accurate ellipse parameters to the coronagraphy than the Source itself.

Regardless from its origin (Reference Object or Source, direct input, moment or PSF), we refer to the final solution as the coronagraphy ellipse. The final set of values found for the coronagraphy ellipse's eccentricity and orientation angle $\theta$ will be used in the coronagraphy of the FOV. The same holds for the ($x_C$, $y_C$) of the Source which hereafter is referred as the ($x_C$, $y_C$) of the coronagraphy centre.

*Appendix A.5. Elliptical rings and apertures*

Rings and apertures are essential in our digital coronagraphy. After sampling the ring pixels – which procedure we describe in this Section – we apply refined quartile statistics to build the Source profile for the coronagraphy of the FOV. Apertures are also important because the pixels inside are used to determine $(x, y)$ centres.

In older task versions, circular rings/apertures were used. Now we use elliptical ones. This allows to cope with elongated images from observations under moderate optical distortions (color refraction without filters, unfocused images, mirror or lens issues, etc), non-sidereal or bad guiding, or strong winds. Unlike with circular shapes, the identification of pixels inside elliptical rings/apertures demands a more elaborated procedure here described.

Given a nominal ellipse with orientation angle $\theta$ with respect to the $x$ axis, eccentricity $e$ with associated semi-major axis $a$ and semi-minor axis $b$, centred at $(x_C, y_C)$, the ellipse ring shall contain pixels not far from the nominal ellipse by more than half its width $w$. Thus, we must compute the closest distance between pixels and the nominal ellipse to find if a pixel belongs to the ellipse ring. This can be done by adapting the receipts described in Eberly (2020).

First, we get new pixel coordinates after shifting the origin to $(x_C, y_C)$ and rotating the axes by $\theta$. The canonical ellipse equation $(X/a)^2 + (Y/b)^2 = 1$ now holds with $a > b$. Then, after taking the pixel coordinate's counterparts $(X_P, Y_P)$ at the first quadrant, one must find the root $\bar{s}$ of $G(s) = 0$ from Eq. A.1:

$$r_0 = \left(\frac{a}{b}\right)^2 , \; z_0 = \frac{X_P}{a} , \; z_1 = \frac{Y_P}{b} ,$$
$$G(s) = \left(\frac{r_0 z_0}{s + r_0}\right)^2 + \left(\frac{z_1}{s + 1}\right)^2 - 1 \quad \text{(A.1)}$$

The root $\bar{s}$ between $-1 + z_1$ and $-1 + \sqrt{r_0^2 z_0^2 + z_1^2}$ is computed by a bisection algorithm adapted from Press et al. (1982). We then define the quantity $\bar{t} = \bar{s} \, b^2$. For $\bar{t} < 0$, $\bar{t} = 0$ and $\bar{t} > 0$ the pixel is respectively inside, over or outside the canonical ellipse. The closest distance $d$ between the pixel $(X_P, Y_P)$ and the canonical ellipse at the point $(X_E, Y_E)$ is finally obtained by the formulae in Eq. A.2.

$$Y_P > 0 , X_P > 0 \begin{cases} X_E = \frac{a^2 X_P}{\bar{t} + a^2} \\ Y_E = \frac{b^2 Y_P}{\bar{t} + b^2} \\ d = \sqrt{(X_P - X_E)^2 + (Y_P - Y_E)^2} \end{cases}$$

$$Y_P > 0 , X_P = 0 \begin{cases} X_E = 0 \\ Y_E = b \\ d = |Y_P - Y_E| \end{cases}$$

$$Y_P = 0 , X_P < (a^2 - b^2)/a \begin{cases} X_E = \frac{a^2 X_P}{a^2 - b^2} \\ Y_E = b \sqrt{1 - \left(\frac{X_E}{a}\right)^2} \\ d = \sqrt{(X_P - X_E)^2 + Y_E^2} \end{cases}$$

$$Y_P = 0 , X_P \geq (a^2 - b^2)/a \begin{cases} X_E = a \\ Y_E = 0 \\ d = |X_P - X_E| \end{cases} \quad \text{(A.2)}$$

Since distances are invariant to origin shifts and rotation, $d$ is the closest distance from the pixel to the nominal ellipse in the non-shifted and non-rotated $(x, y)$ frame of the FOV. Pixels with $d \leq w/2$ are inside the ellipse ring. We adopt $w = \sqrt{2}$ pixels as the width of ellipse rings since this corresponds to the diagonal of a pixel. From the $\bar{t}$ value, we find if the pixel is inside, over or outside an elliptical aperture.

*Appendix A.6. Refined quartile statistics for the rings*

In usual quartile statistics, the lower and upper quarter data values are discarded and the 50 percent middle values used to estimate the average and standard deviation of the sample. The assumption is that values in the extremes may contain corrupted information, while the middle probably contains better information for estimating the sample basic statistical properties.

Quartile statistics improve the processing of rings in comparison to histograms. The inner regions of the Source are specially affected since less pixels are available as in outer regions, resulting in poorer histogram statistics, biased mode counts, unrealistic Source profile pixel counts and thus unsatisfactory coronagraphy results. Quartile statistics are much more robust to less pixels since there are no issues regarding to cell binning.

For extended coronagraphy rings, many outlier pixels have high counts associated to contaminant objects nearby or around the Source. This amount can vary considerably from ring to ring according to the amount of intruding high flux contamination. The same reasoning holds for the lower counts. Since digital coronagraphy is sensible to small changes in the statistics of rings, we now search for optimum lower and upper limits instead of fixing them in one quarter. After sorting ring pixels by counts, the lower and upper limits (e.g. covering 50 percent of the sample) are set as the extreme pixels resulting in the least gradient computed from the pixels' ranks and counts. Better than in normal quartile statistics, these limits encompass pixels with the most typical repetitive values in the ring.

After removal of pixel outliers, we further refine the average count estimate for the ring by the use of two weights.

The distance weight $w_D$ is defined by the distance $d$ of the pixel to the nominal ellipse (see Appendix A.5), as in Eq. A.3. Pixels closer to the nominal ellipse get more weight.

$$w_D = \frac{1}{(1 + d)^2} \quad \text{(A.3)}$$

The mode weight $w_M$ regards to clustering around count values. Pixels with counts closer to each other get more weight. It mimics the search for the peak with histograms, but overcoming fluctuations and small number statistics – a novelty in quartile statistics to the best of our knowledge. The mode weight $w_M$ is computed as in Eq. A.4, where $c$ is the count of the pixel and $c_m$ is the count of each ring pixel $m$ from a total of N ring pixels.

$$w_M = \frac{1}{\sum_{m=1}^{N} [1 + (c - c_m)^2]} \quad \text{(A.4)}$$

The final count average $C$ for a ring of $N$ pixels with counts $c$ (outliers discarded) is computed with Eq. A.5.

$$C = \frac{\sum_{j=1}^{N} w_D(j)\, w_M(j)\, c(j)}{\sum_{k=1}^{N} w_D(k)\, w_M(k)} \quad (A.5)$$

*Appendix A.7. Digital coronagraphy of the FOV*

Here, we describe how the Source profile is obtained pixel by pixel to get the coronagraphed image.

The count values of the Source profile are computed for each pixel $i$ in the coronagraphy working area at their exact $(x_i, y_i)$ coordinates. For each pixel $i$, a nominal ellipse $i$ is constructed. It passes exactly through the $(x_i, y_i)$ coordinates, is centred at $(x_C, y_C)$, has orientation angle $\theta$ and semi-minor and semi-major axis ratios $b/a$ exactly equal to those of the coronagraphy ellipse, i.e. the same eccentricity. For each nominal ellipse $i$, an elliptical ring $r_i$ of width $w$ is set. The pixels inside the elliptical ring $r_i$ are subjected to the quartile statistics described in Appendix A.6. The best count estimate $C$ (Eq. A.5) found for the elliptical ring $r_i$ is then assigned to the pixel $i$ and stored. The procedure is repeated for all pixels $i$ on the coronagraphy working area. The derived brightness Source profile is then stored. The original image is subtracted from this profile to get the digitally coronagraphed image.

Masked bad pixels have values set by the fitted sky background at their $(x_i, y_i)$ coordinates (Appendix A.1). Coronagraphed pixels that eventually get negative count values also have their values reset to the fitted sky background.

The brightness of the sky background in the outer bounds of the working area are matched to that of the surrounding external perimeter. This preserves the photometric properties of all the objects inside and outside of the working area, i.e. the photometric properties of the original images are preserved in the coronagraphy. Working area boundary limits are usually "invisible" in PDC coronagraphed images (see Section 7.1). The setup of working areas is detailed in the PDC's User Guide.

*Appendix A.8. PDC features: multiprocessing, bad pixel masks, outputs*

Digital coronagraphy is computer-time consuming. Since images can be processed independently from each other, PDC allows the processing of multiple images at the same time. It splits the images in separate processes and run them simultaneously. For control, small sub-windows are lunched on screen for inspection of each running process. This is done with native FORTRAN *without* parallelization.

PDC allows for masking bad pixels inside/outside of rectangular or circular regions and ADU ranges. Bad pixels are not processed, avoiding problems e.g. in the sky background flattening and in the coronagraphy itself.

PDC also allows for setting limits to the aperture sizes of the Reference Object and Source. This is useful to avoid aperture contamination by nearby objects.

Besides coronagraphed images, PDC outputs the brightness light profile of the Source. A report is also produced with detailed coronagraphy information, including ellipse parameters and the $(x, y)$ coordinates of the Source, which can be useful in astrometric investigations.

Many more PDC features are detailed in its User Guide.